\documentclass[aip,jcp,reprint,groupedaddress]{revtex4-1}

\usepackage{graphicx}
\usepackage{bm}
\newcommand{\subscript}[1]{\ensuremath{_{\textrm{\footnotesize{#1}}}}}
\newcommand{\ket}[1]{\left| #1 \right\rangle}

\newcommand{\bra}[1]{\left\langle #1 \right|}

\newcommand{\braket}[2]{\left\langle #1 | #2 \right\rangle}
\begin{document}

\title{State-averaged Monte Carlo configuration interaction applied to electronically excited states}

\author{J. P. Coe}
\author{M. J. Paterson}%
\affiliation{ 
Institute of Chemical Sciences, School of Engineering and Physical Sciences, Heriot-Watt University, Edinburgh, EH14 4AS, UK.
}%

\begin{abstract}
We introduce state-averaging into the method of Monte Carlo configuration interaction (SA-MCCI) to allow the stable and efficient calculation of excited states.  We show that excited potential curves for H\subscript{3}, including a crossing with the ground state, can be accurately reproduced using a small fraction of the FCI space.  A recently introduced error measure for potential curves [J. P. Coe and M. J. Paterson, J. Chem. Phys., 137, 204108 (2012)] is shown to also be a fair approach when considering potential curves for multiple states.  We demonstrate that potential curves for LiF using SA-MCCI agree well with the FCI results and the avoided crossing occurs correctly. The seam of conical intersections for CH\subscript{2} found by Yarkony [J. Chem. Phys., 104, 2932 (1996)] is used as a test for SA-MCCI and we compare potential curves from SA-MCCI with FCI results for this system for the first three triplet states.  We then demonstrate the improvement from using SA-MCCI on the dipole of the $2$ $^{1}A_{1}$ state of carbon monoxide. We then look at vertical excitations for small organic molecules up to the size of butadiene where the SA-MCCI energies and oscillator strengths are compared with CASPT2 values [M. Schreiber, M. R. Silva-Junior, S. P. A. Sauer, and W. Thiel, J. Chem. Phys., 128, 134110 (2008)]. We finally see if the SA-MCCI results for these excitation energies can be improved by using MCCIPT2 with approximate natural orbitals when the PT2 space is not onerously large.
\end{abstract}

\maketitle

\section {Introduction}

Monte Carlo configuration interaction (MCCI)\cite{mcciGreer98,mccicodeGreer} randomly augments a configuration space in an iterative scheme where those configurations whose absolute coefficient in the resulting
 solution of the time-independent Schr\"{o}dinger equation in this space is less than a certain amount will eventually be deleted.   This procedure, without prior knowledge of the important configurations or orbitals,  has been demonstrated to produce a compact wavefunction which recreates much of the full configuration interaction (FCI) solution for systems comprising a few atoms. MCCI has recently been used for the calculation of vertical excitations of atoms and small molecules,\cite{GreerMcciSpectra} potential curves of ground states,\cite{MCCIpotentials} electron affinities, ionization energies and multipole moments.\cite{MCCIdipoles}

The main goal of this paper is to investigate whether MCCI with a type of state-averaging (SA-MCCI) can be applied to the calculation of excited states particularly potential curves---including conical intersections
and avoided crossings---using a small fraction of the FCI configurations and without prior knowledge of the important orbitals. 

MCCI represents another approach for the calculation of excited states which may be expected to be particularly suitable for potential curves of small systems. For systems where there is a clearly dominant configuration in the ground-state then single-reference methods based on coupled-cluster such as EOM-CCSD (see, e.g., Ref.~\onlinecite{EOMCCoverview}) may be used to efficiently find excited states, which may be multireference, as long as their main configurations have a substitution level from the ground-state appropriate to the method used.  However along a potential curve, away from the equilibrium geometry, the ground-state wavefunction may have a number of important configurations and here multireference methods such as complete active space self-consistent field (CASSCF),\cite{siegbahn:2384} can be used to capture much of the static correlation perhaps followed by second-order perturbation CASPT2\cite{CASPT2} to recover more of the dynamic correlation.  However to appropriately choose the orbitals for inclusion in the active space requires insight and the calculation can become intractable if the active space is too large. Furthermore the important orbitals may change along the potential curve.  MCCI offers an alternative approach as it depends only on the cut-off value for configurations to be included so in principle, with sufficiently small cut-off and sufficiently long calculation time, it can find compact wavefunctions to describe excited states accurately at all points along a potential curve. Other recent stochastic approaches to excited states include those using projector or diffusion Monte Carlo in Slater determinant space to approach the FCI energy and wavefunction. For example, Ref.~\onlinecite{Nagase10PMC} calculates an excited state by using the projector Monte Carlo results from all the states below to eliminate their component from the initial wavefunction.  Ref.~\onlinecite{FCIQMCexcited} uses $(\hat{H}-S)^{2}$ in the projection operator where $S$ is chosen to be very close to the energy of the state of interest.  Another interesting approach is model space quantum Monte Carlo (MSQMC)\cite{MSQMC} which partitions configuration space and uses projector Monte Carlo to construct an effective Hamiltonian in the smaller space where Slater determinants can move from one space to the other if their importance passes a threshold.  We note that MCCI is not used in this work to find the FCI energy, but rather a very compact wavefunction which captures much of the FCI result yet can be used for further calculations due to its tractable size.   

The calculation of excited state potential curves in MCCI is more challenging than those of the ground state as during an MCCI calculation of, e.g., the first excited state of a given symmetry the method may remove configurations unimportant for the excited state to produce a set of configuration state functions (CSFs) that cannot describe the ground state. In this case the ground state within this set may become similar to the first excited state and the first excited state may be an approximation to a higher excited state.  In the best case this will cause oscillations in the property of interest (e.g. the excited state dipole moment of CO in Ref.~\onlinecite{MCCIdipoles}), but it can cause the diagonalization procedure to fail.  We introduce a type of state-averaging to MCCI (SA-MCCI) to efficiently overcome this problem.   We demonstrate that excited potential curves with conical intersections for H\subscript{3} and CH\subscript{2}, and avoided crossings for LiF can be calculated using SA-MCCI. We then show that this method can prevent the oscillations seen in the MCCI dipole results for the 2 $^{1}A_1$ state of carbon monoxide.\cite{MCCIdipoles}  We finally look at the vertical excitations of organic molecules up to the size of butadiene using SA-MCCI and also consider if the approach of using approximate natural orbitals to construct small SA-MCCI wavefunctions followed by MCCIPT2\cite{MCCInatorb} can more accurately describe these excitation energies.

\section{Method}

We first present a brief recap of the standard MCCI algorithm\cite{mcciGreer98,mccicodeGreer} which usually begins with the CSF constructed from occupied Hartree-Fock (HF) molecular orbitals:

\begin{enumerate}
\item{A set of CSFs is randomly enlarged using symmetry preserving single and double substitutions.}
\item{The Hamiltonian matrix and overlap matrix are constructed and a diagonalization gives the wavefunction coefficients.}
\item{CSFs that have just been added are deleted if the absolute value of their coefficient is less than $c_{\text{min}}$}  
\item{The process continues and every ten iterations all CSFs are considered for deletion (full pruning).}
\end{enumerate}

The integrals of the molecular orbitals are calculated using Molpro\cite{Molpro} for the potential curves in this work, while for oscillator strength calculations of the vertical excitations we use Columbus\cite{Columbus} due to the easy availability of the dipole integrals.  We freeze core orbitals for the calculations of the integrals rather than in the MCCI calculation. This means that fewer calculations are needed in MCCI when constructing matrix elements and fewer integrals need to be stored.  

We use a convergence check when calculating potential curves using MCCI as described in Ref.~\onlinecite{MCCInatorb}.  Here the set of CSFs is not enlarged on the step following a full prune. The maximum difference in the energies after the last three full pruning steps is required to be less than $10^{-3}$ Hartree for the calculation to cease. We employ sixty warmup iterations before the convergence check is implemented and run MCCI on twelve processors.

One approach when working with excited states of a certain symmetry would be to calculate the ground state then ensure that these configurations are always included in an excited-state calculation.  For the vertical excitations of Ref.~\onlinecite{GreerMcciSpectra}, important configurations from the lower states were permanently included\cite{GyorffyThesis} to prevent the collapse of the excited state MCCI calculation.  However for a calculation of the first excited state with a fixed number of iterations this will require twice as many diagonalizations and the size of the excited-state Hamiltonian matrix will always be at least that of the ground state. This problem will clearly become worse as the number of states of interest increases.  We therefore investigate a type of state averaging: in addition to the coefficients for the excited state $s$ of interest  $c_{i,s}$ the coefficients $c_{i,1}$ of the ground state and all other states below $s$ in this symmetry and spin are calculated at each step and when coefficients are considered for removal they are constructed as
\begin{equation}
c_{i}=\sum_{j=1}^{s}|c_{i,j}|.
\end{equation}

This aims to ensure that the set of CSFs found during the MCCI run offers a good description of both the ground and excited states. The size of the wavefunction will be slightly larger as some states that would be removed in a single-state calculation may have a large enough state-averaged value to be included.  However we can start from the HF reference so the states are built up together and the size of the MCCI wavefunction in the early iterations at least is smaller than when using a fixed ground state. Furthermore we reduce the number of diagonalizations by using the Davidson-Liu algorithm\cite{LiuReportNumerical} to calculate the $s$ lowest eigenvalues and eigenvectors simultaneously. The initial guess of expansion vectors $b_{k}$ for the smaller Hamiltonian matrix that is solved in the Davidson algorithm appears to be problematic for excited states but rarely for the ground as the HF configuration will tend to still have a large coefficient after the first addition of configurations in the latter case. Hence, to improve stability of the method, on the first iteration the ground state only is calculated, but new $b_{k}$ vectors from the ground and excited states of the reduced Hamiltonian matrix are created for use in the next iteration.

\section{Results}

\subsection{H\subscript{3} equilateral triangle}

\begin{figure}[ht]\centering
\includegraphics[width=.45\textwidth]{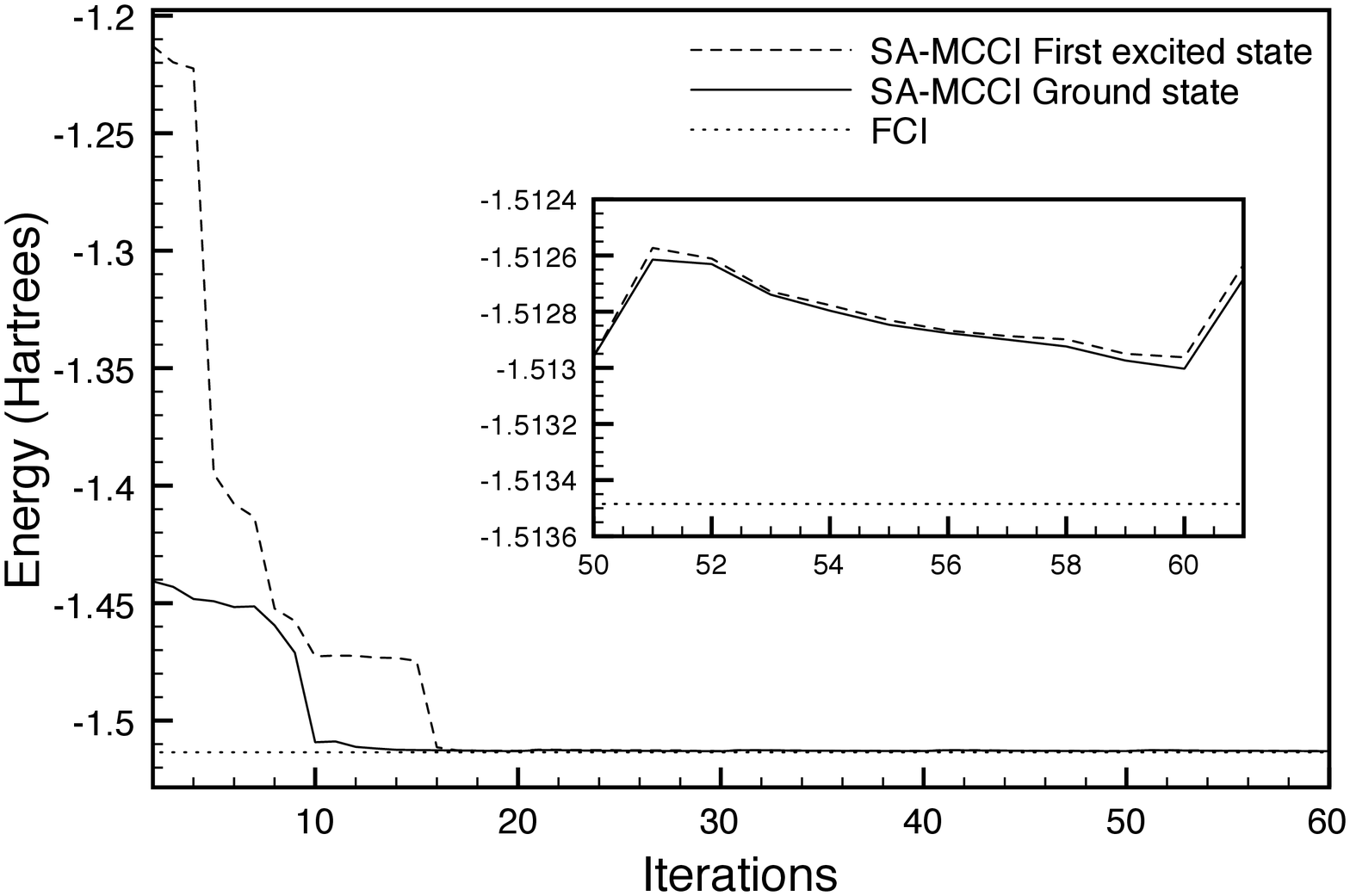}
\caption{Threee hydrogens in an equilateral triangle with a bond length of $2$ angstrom and a cc-pVTZ basis. SA-MCCI results with  $c_{\text{min}}=10^{-3}$ are plotted from iteration 2 as iteration 1 is only the ground state. Inset: Enlargement of the final iterations}\label{fig:H3StateA}
\end{figure}

We test SA-MCCI on three hydrogens in an equilateral triangle using a cc-pVTZ basis and a bond length of $2$ angstrom. At this geometry the ground and first excited state are degenerate and of different symmetry. To test the ability of SA-MCCI to describe excited states at a degeneracy we do not use symmetry.  In this case standard MCCI is not able to calculate the excited state but SA-MCCI, as depicted in Fig.~\ref{fig:H3StateA}, works well.  $352$ CSFs were used to describe both states compared with $36162$ SDs for the FCI when symmetry is not considered. We note that we needed to use symmetry to achieve degeneracy in the Molpro\cite{Molpro} FCI calculations.

\begin{figure}[ht]\centering
\includegraphics[width=.45\textwidth]{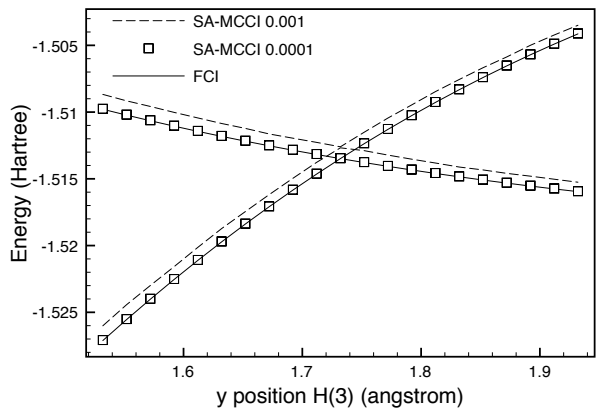}
\caption{SA-MCCI and FCI results for the energy of the ground and first excited state of H\subscript{3} plotted against the $H(3)$ y co-ordinate in angstrom using a cc-pVTZ basis.}\label{fig:EnergiesVH3yvalue}
\end{figure}

Given the following positions in angstrom $H(1)$ $(1,0,0)$, $H(2)$ $(-1,0,0)$ $H(3)$ $(0,\sqrt{3},0)$ we now vary the y position of $H(3)$ to see how SA-MCCI behaves in the vicinity of the point of degeneracy.  Again we do not use symmetry in the SA-MCCI calculation but depict the FCI results of $A_{1}$ and $B_{1}$ symmetry within $C_{2v}$.  Starting from the HF reference appears challenging for SA-MCCI here and often the excited state is much too high in energy.  We attribute this to the configurations comprising the ground-state wavefunction as having zero matrix elements with the excited-state due to symmetry and it seems that   SA-MCCI can sometimes fail to include sufficient states of the excited symmetry when we move away from the point of degeneracy.  To overcome this we use the SA-MCCI states found for the point of degeneracy as the starting point for each calculation.  As we now have sufficient states of both symmetry to begin with then this approach works well:  Fig.~\ref{fig:EnergiesVH3yvalue} shows that with a cut-off of $c_{\text{min}}=10^{-3}$ the shape of the curves near the degeneracy are reproduced.  

 We may quantify the match of this curve using the measure introduced in Ref.~\onlinecite{MCCInatorb}.  There it was shown that the standard deviation of the difference in energies ($\sigma_{\Delta E}$) can be used as a measure of the error between two potential curves. This takes into account all the points and that the curves may be shifted by a constant.  We find $\sigma_{\Delta E}=7.7\times10^{-2}$ kcal/mol for the ground and $\sigma_{\Delta E}=8.6\times10^{-2}$ kcal/mol for the excited curve when the cut-off is $10^{-3}$. While at $c_{\text{min}}=10^{-4}$ the curve is almost indistinguishable from the FCI here $\sigma_{\Delta E}=9.7\times10^{-4}$ kcal/mol for the ground and  $\sigma_{\Delta E}=1.4\times10^{-3}$ kcal/mol for the excited state.  This measure is useful if one wants to quantify how well an approximate method gives the shape of a single curve.  However when we want to quantify how well an approximate method gives the shape of several curves such an approach applied to each curve may not be fair if it results in shifting an excited state curve by a different constant to the ground state.  One way to account for this is to use the vertical excitation energies from the lowest state at each point of the curve, as such a quantity is independent of the overall constant.  However this would not detect if the ground curve was poor but the curves of the second and third state are good. An average of vertical excitation energies between all states could be used to prevent this. However to facilitate easy comparison with results only for the ground-state, we extend our method of Ref.~\onlinecite{MCCInatorb} to multiple states.  Here we average our measure over the states of interest when all the curves from an approximate method may be shifted by a constant $c$

\begin{equation}
A=\frac{1}{s}\sum_{j=1}^{s}\frac{1}{M}\sum_{i=1}^{M}(\Delta E_{i,j} -c)^{2}.
\label{eq:excitedCurvesError}
\end{equation}

Here $\Delta E_{i,j}=E_{i,j}^{FCI}-E_{i,j}^{approx}$  and $E_{i,j}$ corresponds to the energy of state $j$ of at geometry $i$ when we consider $s$ states and $M$ geometries.  We find the minimum of this error with respect to $c$ by setting $\frac{\partial A}{\partial c}=0$.  This results in
\begin{equation}
c=\frac{1}{s} \sum_{j=1}^{s}\frac{1}{M}\sum_{i=1}^{M} \Delta E_{i,j}.
\end{equation}

Putting this into Eq.~\ref{eq:excitedCurvesError} leads to a similar result to Ref.~\onlinecite{MCCInatorb}.
\begin{equation}
\min_{c}A=\sigma^{2}_{\Delta E}
\end{equation}
but now all states of interest, in addition to geometries, are considered for the calculation of the variance. Hence we can use $\sigma_{\Delta E}$ to quantify the match between a number of states and geometries and it will suffice to declare how many states are under consideration.  For the two states near the point of degeneracy depicted in Fig.~\ref{fig:EnergiesVH3yvalue} we find $\sigma_{\Delta E}=8.1\times10^{-2}$ kcal/mol for $c_{\text{min}}=10^{-3}$  and $1.3\times10^{-3}$ kcal/mol for $c_{\text{min}}=10^{-4}$. 

\subsubsection{Excited potential curves and phase change}

\begin{figure}[ht]\centering
\includegraphics[width=.45\textwidth]{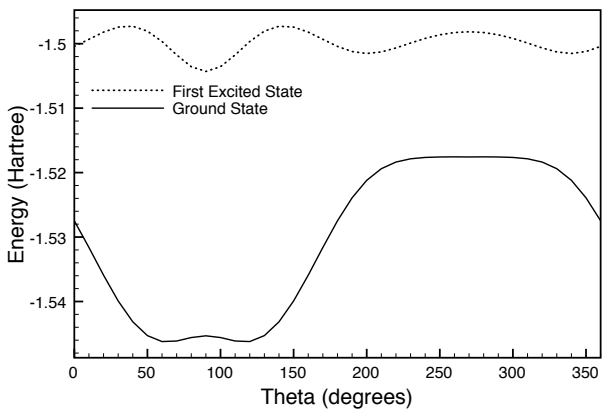}
\caption{SA-MCCI with SDs and $c_{\text{min}}=10^{-4}$ potential curves when moving $H(3)$ around a circle centred at $(0,\sqrt{3},0)$ with radius 0.4 angstrom.}\label{fig:potH3roundCI}
\end{figure}

 For another test of SA-MCCI we consider moving the third hydrogen on a circle of radius $0.4$ angstrom centred on the previous point of degeneracy: $(0,\sqrt{3},0)$ angstrom. We also investigate if SA-MCCI can detect a phase change\cite{LHphase} in this case. We now work with Slater determinants (SDs) rather than CSFs to allow more straightforward calculation of the overlap between SA-MCCI wavefunctions at different geometries. In addition we do not use a convergence check here. Due to the use of SDs we also use a smaller cut-off.  We use the following co-ordinates in angstrom: $H(1)$ $(1,0,0)$, $H(2)$ $(-1,0,0)$ $H(3)$ $(0.4\cos(\theta),\sqrt{3}-0.4\sin(\theta),0)$. Again we do not use symmetry. The energy curve is displayed in Fig.~\ref{fig:potH3roundCI} where we see that no crossings occur. We note that the points at 90 and 270 degrees were not smooth unless the previous point was used as the starting guess.  This again suggests that SA-MCCI has difficulty when including states of a different symmetry when symmetry is not being exploited in the calculation.  We find that the SA-MCCI curves with this small cut-off give a result that is almost the same as that of FCI ($\sigma_{\Delta E}=4.6\times 10^{-3}$) and without symmetry MCCI used 2442 SDs on average compared with the full CI space of 36162 SDs.

  To calculate the overlap between SA-MCCI wavefunctions at different geometries we firstly use Molpro\cite{Molpro} to calculate the overlap of the AOs between different geometries.  The MO coefficients are then used to construct the MO overlaps between the two geometries.  The overlap between SDs $D_{1}$ and $D_{2}$ at different geometries is\cite{Lowdin55}
\begin{equation}
\braket{D_1}{D_2}=\det(\bm{S})
\end{equation}
where $S_{ij}=\int \phi_{1,i}^{*}(\bm{x}) \phi_{2,j}(\bm{x})\bm{dx}$ and $i$ labels the spin-MOs in $D_{1}$ while $j$ labels those in $D_{2}$.  As all coefficients are taken to be real, we find that with a step size of 10 degrees the absolute overlap between consecutive SA-MCCI wavefunctions is always greater than $0.988$ and by requiring the overlap to be positive we see that the wavefunction changes sign when we traverse the circle. The overlap seems to be large enough to suggest that SA-MCCI with SDs can display the expected phase change when traversing around the conical intersection in this case.

We next test SA-MCCI on systems that have a degeneracy or avoided crossing that is not a consequence of the symmetries of the molecule. 

\subsection{LiF avoided crossing}

The adiabatic potential curves for the first two $^{1}\Sigma^{+}$ states of LiF are known to exhibit an avoided crossing, this was seen, for example, by C. W. Bauschlicher Jr. and  S. R. Langhoff using FCI and SA-MRCI.\cite{Bauschlicher88}  There a change from ionic to covalent character was demonstrated for the ground state using the dipole moment while the opposite behaviour was seen for the excited state.  It was later shown\cite{Malrieu95} that CASPT2 could have difficulties with the curve of LiF: a double crossing was observed in this case rather than an avoided crossing. Later, Multi-state CASPT2,\cite{Finley98} which uses multiple SA-CASSCF states as the reference, was demonstrated to be able to describe the curve.

\begin{figure}[ht]\centering
\includegraphics[width=.45\textwidth]{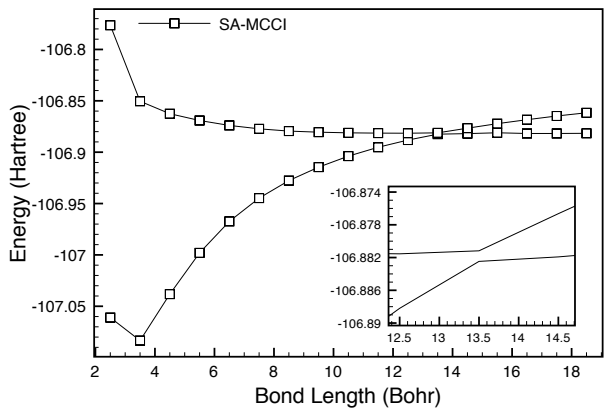}
\caption{LiF ground and first excited singlet state energy of symmetry $A_{1}$ against bond length (Bohr) using SA-MCCI with $c_{min}=3\times10^{-4}$ and the 6-31++G* basis. Inset: enlarged view of the avoided crossing.}\label{fig:LiF631}
\end{figure}

 FCI curves for LiF were also considered in Ref.~\onlinecite{VarandasLiF09} including those for the 6-31++G* basis. Using this basis and three frozen cores we also calculate the FCI curves for a range of geometries with Molpro.\cite{Molpro} The FCI space contains around $7\times 10^{6}$ SDs when considering symmetry.  With FCI the gap at the avoided crossing at $13.5$ Bohr is $0.036$ eV. We return to using SA-MCCI with CSFs and a convergence check. We note that with $c_{\text{min}}=5\times10^{-4}$ the curve at large $R$ was not as smooth as we would have liked as small undulations occurred after $R=15$ Bohr.  However the avoided crossing gap occurred at $13.5$ Bohr, in agreement with the FCI result, and was $0.053$ eV. The results are displayed (Fig.~\ref{fig:LiF631}) for $c_{\text{min}}=3\times10^{-4}$ which used $4186$ CSFs on average and resulted in a gap of $0.035$ eV at $13.5$ Bohr. For $c_{\text{min}}=3\times10^{-4}$ we find that for the two states considered $\sigma_{\Delta E}=0.325$ kcal/mol while this error was almost doubled, although still not too large, at $\sigma_{\Delta E}=0.602$ kcal/mol for the larger cut-off of $c_{\text{min}}=5\times10^{-4}$.

Ref.~\onlinecite{Nagase10PMC} compared projector Monte Carlo FCI with SDs (PMC-SD) for excited states with standard FCI using the basis of Ref.~\onlinecite{Bauschlicher88} (Li (9s 5p)/[4s 2p]; F (9s 6p 1d)/[4s 3p 1d]).  They found that PMC-SD almost gave the same energies as FCI when using around $5\times 10^{6}$ walkers.  The FCI results are reported to 3 decimal places only in the paper so we recreate their FCI results using Molpro.\cite{Molpro}  Two frozen cores are employed which means there are around $7.7\times 10^{7}$ SDs in the FCI when considering symmetry while the mean number of CSFs at $c_{\text{min}}=5\times10^{-4}$ to describe both the ground and excited state in SA-MCCI was $2734$.  

\begin{figure}[ht]\centering
\includegraphics[width=.45\textwidth]{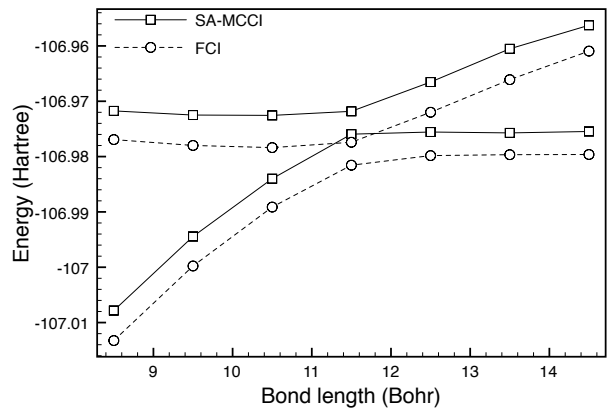}
\caption{Potential curves for the LiF ground and first excited state of symmetry $A_{1}$ against bond length (Bohr) calculated with FCI and SA-MCCI at $c_{\text{min}}=5\times10^{-4}$ when using the basis of Ref.~\onlinecite{Bauschlicher88}.}\label{fig:LiFplot1}
\end{figure}

We see in Fig.~\ref{fig:LiFplot1} that, although the SA-MCCI results are noticeably higher in energy than the FCI, the avoided crossing is reproduced and the shapes of the curves appear close to that of the FCI.  The avoided crossing gap at $11.5$ Bohr is $0.113$ eV when using FCI while SA-MCCI gives $0.112$ eV.  We find for the potential curves from the two SA-MCCI states that $\sigma_{\Delta E}=0.368$ kcal/mol.  When using both the PMC-SD and FCI results to 3 decimal places as reported in  Ref.~\onlinecite{Nagase10PMC} we see that $\sigma_{\Delta E}=0.312$ kcal/mol.  If we also restrict the SA-MCCI comparison to three decimal places then we find that the error in the SA-MCCI curves increases slightly to $\sigma_{\Delta E}=0.372$ kcal/mol. Although the energies are higher in SA-MCCI the shape of the curves are very good with only a slightly larger error than PMC-SD despite using a wavefunction consisting only of a few thousand CSFs. However, given the paucity of points and that the $\sigma_{\Delta E}$ value for PMC-SD is based on results rounded to three decimal places then caution should be used with the comparison.

\subsubsection{Computational efficiency}

In Ref.~\onlinecite{mcciGreer95} the scaling of MCCI with Slater determinants was considered.  The cost of iterative diagonalization was taken as $O(N^{2})$ where $N$ is the number of configurations so a possible reduction of time when using MCCI is $O(N_{MCCI}^{2}/N_{FCI}^{2})$.  However the number of iterations needed for MCCI may affect the scaling and it is noted that FCI programs can improve efficiency by exploiting the sparsity of the Hamiltonian matrix.  
  
Comparing SA-MCCI directly to methods other than FCI is problematic because SA-MCCI can be applied to single-reference or multi-reference wavefunctions and ground states or excited states with or without degeneracies.  Furthermore SA-MCCI produces a compact wavefunction that can be used in the subsequent calculation of properties. Efficient methods for specific problems such as single-reference wavefunctions would be expected to surpass MCCI in their domain of applicability but may be qualitatively wrong for other problems.
CASSCF perhaps followed by perturbation or MRCI can be applied generally but its efficiency is dependent on the choice of active space and the time required to find an appropriate active space, which may be difficult for excited states, and it is not clear how to take these aspects into account. The suggestion of an appropriate active space for excited states could perhaps be a future application of MCCI with large cut-offs.

A comparison with FCI is still not straightforward as FCI memory requirements increase quickly with the size of the basis and so can prevent FCI calculations when MCCI can still be run.  If FCI is possible then  the MOLPRO\cite{Molpro} FCI program used here scales as\cite{MolproFCI1} $O(N_{FCI}m^{4})$ where $m$ is the number of orbitals. Hence if the fraction of FCI space needed by MCCI is sufficiently small then MCCI will be more efficient than FCI, but we cannot know beforehand if this is this case. Furthermore the cut-off value and convergence criteria for the MCCI calculation will affect the balance between time and accuracy.  With these caveats we briefly look at the computational cost of the results for LiF as they capture the behaviour of the FCI results, demonstrated by the small errors ($\sigma_{\Delta E}$), and the calculations are of reasonable duration.  For the 6-31++G* basis the FCI calculation required 0.43 hours per point on average while the SA-MCCI calculations used 0.09 hours per point and 0.21 hours per point for cut-offs of $5\times10^{-4}$ and $3\times10^{-4}$ respectively.  However the SA-MCCI calculation was run on $12$ processors while the FCI program is not parallelized.  SA-MCCI therefore requires 1.12 processor hours per point for the large cut-off and 2.52, for the smaller cut-off.  For the results when using the basis of Ref.~\onlinecite{Bauschlicher88} the number of FCI configurations increases by an order of magnitude and now the FCI calculation uses 5.5 hours per point and the SA-MCCI calculation for $c_{\text{min}}=5\times10^{-4}$ required 0.09 hours per point or 1.11 processor hours per point.  Although when considering processor hours this implementation of SA-MCCI was less efficient than FCI for this system when the FCI space consisted of $\sim 7\times 10^{6}$ SDs, it appeared more efficient when the FCI space was $7.7\times 10^{7}$ however we acknowledge that the larger calculation was over a smaller range of geometries.  These values approximately conform to the expected scalings in that the FCI space has increased by an order of magnitude and so has the FCI calculation time while the size of the SA-MCCI space is similar resulting in similar calculation time for the SA-MCCI calculation.  We note that if the size of the FCI space were to continue to increase in magnitude then the FCI calculation would quickly become intractable and the small number of configurations used for the SA-MCCI wavefunctions means that there can be further efficiencies, compared with the FCI wavefunction, if they are later used for the calculation of properties.

\subsection{CH\subscript{2} conical intersections}

\begin{figure}[ht]\centering
\includegraphics[width=.45\textwidth]{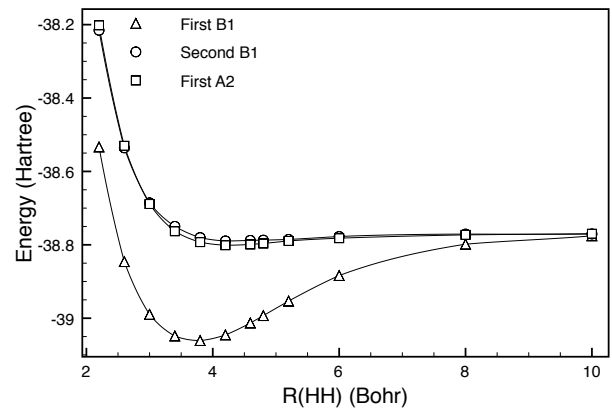}
\caption{Energy of the $1 ^{3}A_{2}$ and $2 ^{3}B_{1}$ states of CH\subscript{2} when using the basis of Ref.~\onlinecite{YarkonyCH2} and one frozen orbital plotted against R(HH) Bohr. Here $\theta=90$ and R(CH) is not displayed.}\label{fig:CH2YarkonyBasis}
\end{figure}

Conical intersections of the $2$ $^{3}A''$ and $3$ $^{3}A''$ states of CH\subscript{2} when using $C_{s}$ symmetry ( $1$ $^{3}A_{2}$ and $2$ $^{3}B_{1}$ when the molecule has $C_{2v}$ symmetry)  were found directly in Ref.~\onlinecite{YarkonyCH2}.  When the angle of the carbon atom from the centre of mass of the hydrogens was at $\theta=90$ degrees a seam of conical intersection was reported.  We test if by using the basis of Ref.~\onlinecite{YarkonyCH2} SA-MCCI can give degenerate energies when following the seam found in Ref.~\onlinecite{YarkonyCH2}. Here the carbon to hydrogen distance R(CH) and hydrogen to hydrogen distance R(HH) are both varied.   We see in Fig.~\ref{fig:CH2YarkonyBasis}  that the shape of the curves are as expected and that the $1$ $^{3}A_{2}$ and $2$ $^{3}B_{1}$ states from SA-MCCI are essentially degenerate for much of the curve although there is a small gap noticeable around $R(HH)=4$ Bohr.  However this only used in the region of $10^{4}$ CSFs (12718 for the $2$ $^{3}B_{1}$ state on average) compared to around $5\times 10^{8}$ SDs that would be needed for a FCI in $C_{2V}$ symmetry.  We note that the second-order CSF expansions using SA-CASSCF orbitals used to calculate the curve in Ref.~\onlinecite{YarkonyCH2} required around $6\times 10^{5}$ configurations. It is not clear whether this small gap is due to SA-MCCI not being accurate enough here or that the FCI would also display a gap due to the FCI conical intersection seam for $\theta=90$ degrees occurring at slightly different geometries to the results of Ref.~\onlinecite{YarkonyCH2}.
 
We now look at potential curves when $\theta \neq 90$ degrees, i.e., $C_{s}$ symmetry with a smaller basis so that we may compare with FCI results.  Using a cc-pVDZ basis with one frozen core and $c_{\text{min}}=5\times 10^{-4}$ we fix the separation of the hydrogens at $2$ angstrom and vary the angle and distance of the carbon atom from the centre of mass of the hydrogens.

\begin{figure}[ht]\centering
\includegraphics[width=.45\textwidth]{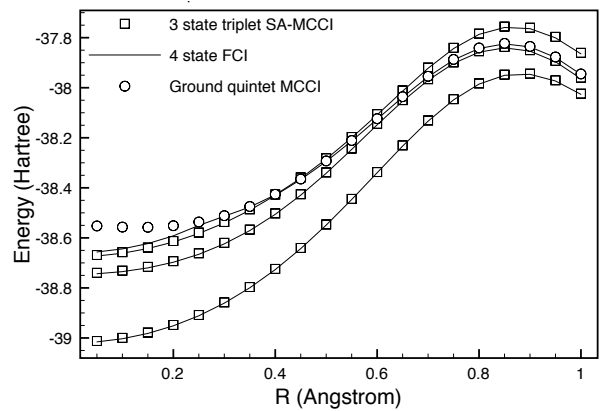}
\caption{SA-MCCI at $c_{\text{min}}=5\times 10^{-4}$ and FCI results for the A'' states of CH\subscript{2} at $\theta=30$ degrees when using the cc-pVDZ basis.}\label{fig:CH2vdztheta30}
\end{figure}

The assignment of the curves in Fig.~\ref{fig:CH2vdztheta30} is not straightforward  when $\theta=30$ degrees as in the FCI triplet calculation the fourth state crosses the third around $R=0.45$  and this state is not found in the triplet SA-MCCI calculation.  Due to the use of CSFs in MCCI we can calculate the ground state quintet and find that this corresponds to most of the FCI curve which was not detected in the SA-MCCI triplet calculation.  The occurrence of this state is attributed to the use of Slater determinants in a FCI calculation resulting in the possible inclusion of higher spin states.  We see that there is not a crossing of the 2 $^{3}A''$ and 3 $^{3}A''$ states for the range considered.   Despite the complexity of the potential energy surface we find that SA-MCCI performs well here: for the first three triplet states we find that $\sigma_{\Delta E}=0.34$ kcal/mol.

\begin{figure}[ht]\centering
\includegraphics[width=.45\textwidth]{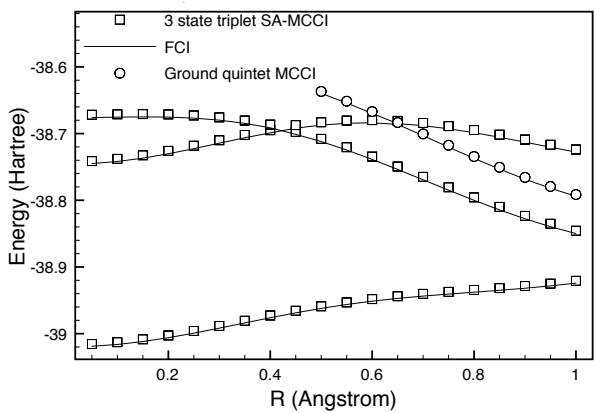}
\caption{SA-MCCI at $c_{\text{min}}=5\times 10^{-4}$ and FCI results for the A'' excited states of CH\subscript{2} at $\theta=60$ degrees when using the cc-pVDZ basis.}\label{fig:CH2vdztheta60}
\end{figure}

At $\theta=60$ degrees there is a conical intersection between the second and third state at $R\approx 0.45$ angstroms and this is described well by SA-MCCI (Fig.~\ref{fig:CH2vdztheta60}).  There is again a crossing between the third and fourth FCI states which is not detected as a triplet using SA-MCCI and as before we find that it is due to a ground-state quintet.  When using the first three triplet states we find a good match between SA-MCCI and FCI with $\sigma_{\Delta E}=0.28$ kcal/mol.  

 The FCI calculation has an SD space of around $1.1\times 10^{6}$ when symmetry is taken into account while SA-MCCI with $c_{\text{min}}=5\times 10^{-4}$ required, on average, $7247$ CSFs for $\theta=60$ degrees and $6425$ CSFs for $\theta=30$ degrees.  With this very small fraction of the FCI space SA-MCCI was still able to produce accurate excited potential curves and demonstrate conical intersections between the second and third triplet states. 

\subsection{Dipole of the $2$ $^{1}A_1$ state of CO}

\begin{figure}[ht]\centering
\includegraphics[width=.45\textwidth]{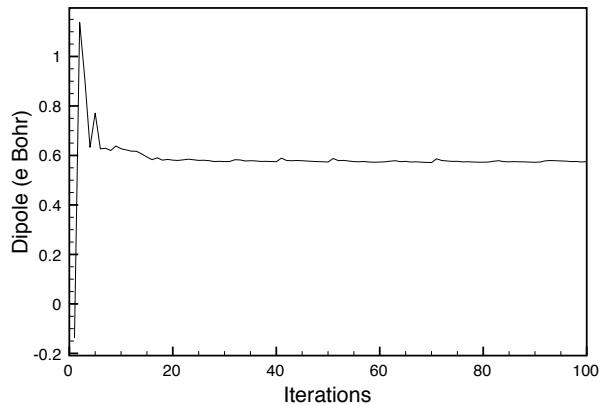}
\caption{MCCI results for the dipole moment (e Bohr) against iteration number for the $2$ $^{1}A_{1}$ state within $C_{2v}$ symmetry of CO at $2.116$ Bohr when using the cc-pVDZ basis set and $c_{\text{min}}=10^{-3}$.}\label{fig:CoExcitedipoleVDZ}
\end{figure}

As alluded to in the introduction, in earlier work\cite{MCCIdipoles} it was found that the dipole of the first excited state of  $A_{1}$ symmetry for carbon monoxide could be calculated with MCCI and was close to the FCI result, but there were large oscillations in the MCCI dipole during the calculation.  These oscillations caused the dipole to fluctuate from around $0.2$ e Bohr to almost $0.8$ e Bohr. We attribute this to the MCCI set of CSFs at times being unable to describe the ground state so the first excited state is now a higher excited state.  We now check if SA-MCCI can remedy this.  The SA-MCCI dipole of the first excited singlet state of CO is displayed in Fig.~\ref{fig:CoExcitedipoleVDZ}.  On iteration $100$ the state averaging type calculation used $18765$ CSFs while  the standard calculation had $8988$ CSFs. Here the standard MCCI result was $0.61$ e Bohr while the SA-MCCI result was $0.58$ e Bohr which is slightly closer to the FCI result of  $0.56$ e Bohr.\cite{MCCIdipoles} The erratic behaviour when not using state-averaging has been eliminated in the SA-MCCI case.  The larger space means that more time is needed for the SA-MCCI calculation here but the lack of oscillations means that convergence of the calculation could be checked to finish the computation in much fewer iterations.  

\subsection{Vertical excitations of organic molecules}

Electronic excitation energies of medium-sized organic molecules were calculated using CASPT2 and a range of coupled-cluster approaches in Ref.~\onlinecite{OrganicBenchmarks} when using a def1-TZVP basis.\cite{def1TZVPbasisRef} Vertical excitation energies using a geometry optimized using MP2/6-31G* were reported.  FCI results are not available due to the very large size of the FCI space for these molecules when using this basis.   We therefore compare MCCI results with these benchmarks for some of the smaller organic molecules. As the molecules in the ground state are in, or close to, their def1-TZVP equilibrium structure then one would expect methods based on a single reference to be appropriate except when the excited state has important configurations involving substitutions beyond the method employed.  For example EOM-CCSD performs well for many of the systems in Ref.~\onlinecite{OrganicBenchmarks} but for the 2 $^{1} A'$ transition of formamide it gives 4.52 eV compared with 7.44eV for CASPT2.  These systems therefore represent a useful benchmark of MCCI even if MCCI would not be the first choice for the calculation of some of these states.  Particularly of interest here are those excitations within a symmetry class to allow the efficacy of SA-MCCI to be observed.

For fairness, excitations energies to states of the same symmetry use the energy from the SA-MCCI calculation for the ground and excited states. While excitations between different symmetries use the energies from single state calculations unless otherwise stated. Core orbitals were frozen for each molecule: two for ethene; two for formaldehyde, three for cyclopropene, three for formamide and four for E-butadiene. Calculations were carried out at $c_{\text{min}}=10^{-3}$ then the resulting converged wavefunction was used as the starting point for a $c_{\text{min}}=5\times 10^{-4}$ calculation.

 In preliminary calculations we saw that the excitation energy to the $1$ $^{1}B_{1}$ state of formaldehyde appeared anomalously high given that the size of the FCI space is much less than cyclopropene and comparable with ethene.  We started this calculation with a $B_{1}$ configuration formed by exchanging the HOMO of $B_{1}$ symmetry with the LUMO of $A_{1}$ symmetry in the HF reference.  We now label this configuration $CSF_{1}$. MCCI should be insensitive to the choice of reference configuration as long as the symmetry is correct and the cut-off is low enough.  It seems that the latter condition was not satisfied here as a CIS calculation suggests that the configuration formed by swapping the highest occupied $A_{1}$ orbital with the lowest unoccupied $B_{1}$ orbital in the HF reference ($CSF_{2}$) is more important based on its coefficient.  

To reduce the chance of the reference affecting the result and reduce the input needed, we modified MCCI so that on the first iteration all single substitutions of the HF reference that give a required symmetry are used to create the initial MCCI space.

In this case we get a lower excitation energy, and the same dominant configuration ($CSF_{2}$) as suggested by CIS, at the end of the MCCI run.  We find that when starting with $CSF_{1}$ then this initial configuration remains as the most important although $CSF_{2}$ appears in the final wavefunction with a small coefficient ($\sim 10^{-2}$) in a $c_{\text{min}}=0.001$ calculation. An EOM-CCSD calculation gives the second excited state of $B_{1}$ symmetry with a dominant excitation which is the HOMO $B_{1}$ to the LUMO $A_{1}$.   This suggests that MCCI has become confined in the second excited state when starting with $CSF_{1}$ and that the other CSFs, at this cut-off, are such that they appear to support $CSF_{1}$ so that it remains dominant even when $CSF_{2}$ is included.  

We use the approach of including all single-substitutions resulting in a required symmetry from the HF reference as the starting point for all subsequent calculations except for multiple states of the completely symmetric representation when this approach caused problems in the diagonalization. 

In table \ref{tbl: Excitation EnergiesDef1} we display the singlet excitation energies for organic molecules in order of increasing FCI space (table \ref{tbl:Fraction of SD spaceDef1}).    The formaldehyde results of different symmetry are close to the CASPT2 results at a cut-off of $10^{-3}$ and become closer as $c_{\text{min}}$ is lowered to $5\times10^{-3}$. The SA-MCCI result for the $2$ $^{1} A_{1}$ excitation is fairly large compared with the CASPT2 result, but is still within 0.9eV of it. For ethene the results agree very well with CASPT2 with the smallest cut-off MCCI excitation energy a little bit lower.  The difference between the methods is more noticeable for cyclopropene where the $c_{\text{min}}=10^{-3}$ results are larger by more than 1eV although the ordering of the states agrees. Reducing the cut-off reduces the difference to about 0.7eV.  The $0.001$ cut-off result for the 1 $^{1}A''$ state of formamide agrees surprisingly well with the CASPT2 result but this may be attributed to the ground state energy still being somewhat high when taken from the SA-MCCI 3 A' calculation.  The $2$ $^{1}A'$ state excitation is around 1.4eV higher than the CASPT2 result at the lower cut-off but, interestingly, the $3$ $^{1}A'$ state excitation is lower at both cut-offs.  E-butadiene had a similar problem with the larger cut-off giving essentially the HF  wavefunction for the $1$ $^{1}A_{g}$ state and the use of the ground state energy of the  $2$ $^{1}A_{g}$ in its place leading to a better match for the 1 $^{1}B_{u}$ excitation at this cut-off with CASPT2 than at the smaller cut-off.  However they are both more than 1 eV higher. The agreement with CASPT2 is poorest for 2 $^{1}A_{g}$ here with a 3 eV difference even at $c_{\text{min}}=5\times10^{-4}$ suggesting that the fraction of the FCI space considered by MCCI is too small here. The 2 $^{1} A_{g}$ state is considered a doubly excited state\cite{BuAnd2AgPolyenes} and we see in Ref.~\onlinecite{OrganicBenchmarks} that EOM-CCSD results in a value of 7.42 eV.  Table \ref{tbl:Fraction of SD spaceDef1} confirms that not only does e-butadiene have the largest FCI space but the percentage used in the MCCI wavefunction is substantially smaller than the other systems considered.  This suggests the use of second-order perturbation on the MCCI wavefunction\cite{MCCInatorb} (MCCIPT2) to see if we can account for some of the large number of configurations that are neglected and improve the excitation energies.  Firstly, however, we will calculate the oscillator strength for the transitions considered and compare this with the CASPT2 results.

\begin{table}[t]
\centering
\caption{MCCI excitation energies (eV) compared with CASPT2 using the Def1-TZVP basis.} \label{tbl: Excitation EnergiesDef1}
\begin{tabular*}{8.5cm}{@{\extracolsep{\fill}}lcccc}
\hline
\hline
Molecule & Excitation &  0.001  & 0.0005 & CASPT2\cite{OrganicBenchmarks}    \\
\hline

Formaldehyde & 1 $^{1}A_{2}$ & 4.45  & 4.17   & 3.98 \\
 & 1 $^{1}B_{1}$ & 9.69  & 9.40  &  9.14 \\
 & 2 $^{1}A_{1}$ & 10.54  & 10.14  & 9.31 \\

 Ethene & 1 $^{1}B_{1u}$ & 8.73  & 8.57   & 8.62 \\

Cyclopropene & 1 $^{1}B_{1}$ & 8.02  & 7.44   & 6.76\\
 & 1 $^{1}B_{2}$ & 8.30  & 7.76  & 7.06 \\

Formamide\footnote{0.001 Ground state from the SA-MCCI $3$ $^{1}A'$ calculation as the $1$ $^{1}A'$ calculation gave essentially HF energy.} & 1 $^{1}A''$ & 5.39 & 6.70 & 5.63 \\
 & 2 $^{1}A'$ & 9.11 & 8.85  &  7.44 \\
 & 3 $^{1}A'$ & 10.18 & 9.93 & 10.54 \\

E-Butadiene\footnote{0.001 Ground state from the SA-MCCI $2$ $^{1}A_{g}$ calculation as the 1 $^{1}A_{g}$ calculation gave essentially the HF energy.} & 1 $^{1}B_{u}$ & 7.58   & 7.79   & 6.47 \\
& 2 $^{1}A_{g}$ & 10.33 & 9.74   & 6.83 \\
\hline
\hline
\end{tabular*}
\end{table}

\begin{table}[t]
\centering
\caption{def1-TZVP Ground state fraction when using $5\times10^{-4}$ } \label{tbl:Fraction of SD spaceDef1}

\begin{tabular*}{8.5cm}{@{\extracolsep{\fill}}lcc}
\hline
\hline
Molecule & FCI SD space &  Fraction MCCI CSFs    \\
\hline
Formaldehyde & $3.8\times10^{13}$  & $2\times 10^{-8}\%$  \\
Ethene & $3.1\times10^{14}$   &  $3\times 10^{-9}\%$  \\
Cyclopropene & $1.4\times 10^{20}$   & $2\times 10^{-14}\%$  \\
Formamide & $3.6\times 10^{21}$ & $10^{-15}\%$  \\
E-Butadiene & $3 \times 10^{28}$   & $2\times 10^{-22}\%$ \\

\hline
\hline
\end{tabular*}
\end{table}

\subsubsection{Oscillator strengths}
When using atomic units, the electric dipole oscillator strength may be written

\begin{equation}
f_{ab}=\frac{2}{3} \Delta E | \bm{D}_{ab}|^{2}
\end{equation}
where
\begin{equation}
 \bm{D}_{ab}=\bra{\Psi_{a}} \bm{\hat{r} }\ket{\Psi_{b}}.
\end{equation}

As we freeze orbitals before producing the integrals then this needs to taken into account in the transition dipole moment calculation by using 
\begin{equation}
\bm{D}_{ab}=\bm{r_{core}}\braket{\Psi_{a}}{\Psi_{b}}+\bra{\Psi_{a}}\bm{\hat{r}}\ket{\Psi_{b}}.
\end{equation}

The oscillator strength would perhaps be expected to be more sensitive to the details of the wavefunction than the excitation energy. The results are displayed in table \ref{tbl:fvaluesDef1} and the absolute differences between MCCI and CASPT2 oscillator strengths are generally reasonably small. Although the percentage errors would be much higher than those of the excitation energies in many cases.  The MCCI formaldehyde f values are similar in order of magnitude to those of CASPT2.  The 1 $^{1}B_{1}$ value for MCCI is around ten times less than that of CASPT2 but both results are very small. Given the agreement with the excitation energies the MCCI f value at around 1 being more than double that of CASPT2 for the 2 $^{1}A_{1}$ transition is unexpected.  The methods however agree well for ethene and also for cyclopropene although the very small 1 $^{1}B_{1}$ values would have a large percentage error in the latter case. The 1 $^{1}A''$ results for formamide are small but agree well, however the result for 2 $^{1}A'$ is very different, when using MCCI at the smaller cut-off, to CASPT2. There is also a discrepancy for the 3 $^{1}A'$  result but this is less marked.  The e-butadiene SA-MCCI results are also fairly good agreement with CASPT2.

\begin{table}[t]
\centering
\caption{MCCI and CASPT2 oscillator strengths using the def1-TZVP basis.} \label{tbl:fvaluesDef1}
\begin{tabular*}{8.5cm}{@{\extracolsep{\fill}}lcccc}
\hline
\hline
Molecule & Excitation &  0.001  & 0.0005 & CASPT2\cite{OrganicBenchmarks}    \\
\hline

Formaldehyde & 1 $^{1}A_{2}$ & 0  & 0  & - \\
 & 1 $^{1}B_{1}$ & $1.60\times10^{-3}$  & $1.78\times10^{-3}$    &  $1.3\times10^{-2}$ \\
 & 2 $^{1}A_{1}$ & 1.12  & 1.07  & 0.451 \\

 Ethene & 1 $^{1}B_{1u}$ & 0.494  & 0.443   & 0.513 \\

Cyclopropene & 1 $^{1}B_{1}$ & $3.11\times10^{-3}$  & $3.03\times10^{-3}$   & $10^{-2}$\\
 & 1 $^{1}B_{2}$ & 0.242  & 0.224   & 0.234 \\

Formamide\footnote{0.001 Ground state from the SA-MCCI $3$ $^{1}A'$ calculation as the $1$ $^{1}A'$ calculation gave essentially HF energy.} & 1 $^{1}A''$ & $1.76\times10^{-3}$ &  $2.3\times10^{-3}$ & $10^{-3}$ \\
 & 2 $^{1}A'$ & $0.429$ & $9.34\times 10^{-2}$  &  0.479 \\
 & 3 $^{1}A'$ & 0.632 & $0.441$ & 0.163 \\

E-Butadiene\footnote{0.001 Ground state from the SA-MCCI $2$ $^{1}A_{g}$ calculation as the 1 $^{1}A_{g}$ calculation gave essentially the HF energy.} & 1 $^{1}B_{u}$ &  1.07 & 1.19   & 0.783 \\
& 2 $^{1}A_{g}$ & 0 & 0   & - \\
\hline
\hline
\end{tabular*}
\end{table}

\subsubsection{MCCIPT2 results with approximate natural orbitals}

Second-order perturbation theory on a MCCI wavefunction (MCCIPT2)\cite{MCCInatorb} was shown to improve the accuracy of the N\subscript{2} potential curve in Ref.~\onlinecite{MCCInatorb} and approximate natural orbitals were demonstrated to be able to increase the efficiency of a MCCI calculation.  We now investigate how MCCIPT2 with approximate natural orbitals fares in the calculation of the excited states.   MCCIPT2 adapts a CI perturbation scheme from Ref.~\onlinecite{HarrisonFCIperturbation} to work with MCCI and employs an efficient approach for the  checking and removal of duplicate CSFs in the PT2 space.  Here any configurations that contribute more than a certain value to the perturbative energy estimate are added to the MCCI CSF space and a new MCCI wavefunction is found. This should mean that intruder states pose less of a problem when calculating the PT2 correction to excited state energies in this method. The PT2 part of the calculation is currently run in serial for CSFs.

We require a consistent cut-off and also the MCCIPT2 calculation to contain around $10^{8}$ CSFs or fewer so that the calculation could be completed in a reasonable time with the current version of the code.  To this end we used a reasonably large cut-off of $c_{\text{min}}=5\times10^{-3}$ for the MCCI calculations.  We hypothesise that sufficient important configurations should be found, but that the MCCI wavefunction is compact enough so that the PT2 calculation is not excessively challenging when the FCI space becomes extremely large.  To try to improve the efficiency of the calculation we used approximate natural orbitals (NOs) where QCISD\cite{QCISD} NOs are employed for the ground-state while MCCI NOs\cite{MCCInatorb} are used for other calculations. The MCCI NOs are calculated using Slater determinants (SDs). When using a starting configuration consisting of all single substitutions resulting in a required symmetry from the HF wavefunction the Davidson-Liu algorithm had problems diagonalizing the Hamiltonian matrix on the first step. In this case we carried out a complete diagonalization of the Hamiltonian matrix and the eigenvectors of interest were then used to create the initial $b_{k}$ vectors for the Davidson-Liu routine.

 Where excited states of the same symmetry were used SA-MCCI was utilised and the state average of the first-order reduced density matrix $\bm{\gamma}=\frac{1}{s}\sum_{s} \bm{\gamma}_{s}$ was used to attempt to fairly construct MCCI NOs.  The procedure for these calculations is summarised below:

\begin{enumerate}
\item{Calculation of SA-MCCI NOs using SDs with 50 iterations and $c_{\text{min}}=5\times10^{-4}$ with a starting configuration comprising all symmetry consistent single substitutions  for all but ground state calculations where the QCISD NOs were calculated.}
\item{SA-MCCI with NOs, a starting configuration comprising all symmetry consistent single substitutions and $c_{\text{min}}=5\times10^{-3}$ until convergence in the energy of  $0.001$ Hartree.}  
\item{MCCIPT2 is then carried out on the converged wavefunction.  Any CSF in the PT2 space that contributes more than the cutoff is incorporated and the process repeated until this does not occur.}
\end{enumerate}

We see in table~\ref{tbl:Def1TZVPExcitationEnergiesPT2} that the MCCIPT2 results are closer to the CASPT2 results than the MCCI results were.  The PT2 calculations included from one million to a hundred million configurations which tended to increase as the FCI space did.  Now the discrepancy is always within 1.1eV, with formamide giving the largest difference. The formamide 3 $^{1} A'$ excitation is noticeably lower than the CASPT2 result using MCCIPT2 and using SA-MCCI which suggests that the FCI result may be lower than the CASPT2 estimate.  The order of the states is the same except for cyclopropene where the close $1$ $^{1}B_{1}$ and $1$ $^{1}B_{2}$ excitations have swapped round.  We repeated the cyclopropene calculations with $c_{\text{min}}=3\times10^{-3}$ which resulted in a larger MCCI wavefunction and PT2 space: around $10^{8}$ configurations were needed for the ground state and $2\times10^{8}$ for the excited states in the PT2 calculation.  The $1$ $^{1} B_{1}$ transition was again higher in energy at $7.46$ eV but was closer to the $1$ $^{1}B_{2}$ transition occurring at $7.42$ eV than before.     
\begin{table}[t]
\centering
\caption{Excitation energies (eV) from MCCIPT2 with $c_{\text{min}}=5\times10^{-3}$ compared with CASPT2 when using a def1-TZVP basis} \label{tbl:Def1TZVPExcitationEnergiesPT2}
\begin{tabular*}{8.5cm}{@{\extracolsep{\fill}}lcccc}
\hline
\hline
Molecule & Excitation &  MCCIPT2  & CASPT2\cite{OrganicBenchmarks} & PT2 configs \\
\hline

Formaldehyde & 1 $^{1}A_{2}$ &  3.98 & 3.98 & $4.2\times 10 ^{6}$   \\
 & 1 $^{1}B_{1}$ & 9.08 &  9.14  &  $4.7\times 10^{6}$  \\
 & 2 $^{1}A_{1}$   & 10.22   & 9.31 &  $5.8\times 10^{6}$  \\

 Ethene & 1 $^{1}B_{1u}$  & 8.20   & 8.62 & $5.8\times 10^{6}$    \\

Cyclopropene & 1 $^{1}B_{1}$  & 7.35   & 6.76 & $4.3\times 10^{7}$  \\
 & 1 $^{1}B_{2}$  & 6.94   & 7.06 & $3.5\times 10^{7}$ \\

Formamide& 1 $^{1}A''$  & 6.75   & 5.63 & $6.6\times 10^{7}$ \\
 & 2 $^{1}A'$  & 8.53 &  7.44 & $9.5\times 10^{7}$  \\
 & 3 $^{1}A'$  & 9.69 & 10.54 & $9.5\times 10^{7}$ \\

E-Butadiene & 1 $^{1}B_{u}$ & 5.89    & 6.47 & $3.4\times 10^{7}$  \\
& 2 $^{1}A_{g}$ &  7.18  & 6.83 & $1.6\times 10^{8}$  \\
\hline
\hline
\end{tabular*}
\end{table}

We also look at the excitation energies with MCCIPT2 when using the improved and larger def2-TZVP basis sets.\cite{def2TZVPbasisRef}  The results are displayed in table \ref{tbl:Def2TZVPExcitationEnergiesPT2} and we see that the values are similar, particularly the different order of the cyclopropene excitations is again observed.  The excitation energy for the $^{1}B_{u}$ state of e-butadiene is now much higher than with the def2-TZVP basis but we note that both MCCIPT2 results are within 1eV either side of the CASPT2 result in the def1-TZVP basis. This sensitivity suggests that we are perhaps not including enough configurations in the e-butadiene $B_{u}$ calculations at this level of cut-off: the def1-TZVP MCCI wavefunction used for the PT2 calculation comprised 169 configurations for the $B_{u}$ state and 662 for the 2 $A_{g}$ state while when using the def2-TZVP basis the number of CSFs was 160 and 574 respectively.  However this is the most challenging calculation due to the large space: before the removal of duplicate CSFs in the MCCIPT2 calculation there were more than half a billion CSFs for the 2 $^{1}A_{g}$ state of e-butadiene. Hence a smaller cut-off calculation with the current version of the code would be unreasonably time consuming.  We note that the ordering of the e-butadiene states agrees with Ref.~\onlinecite{OrganicBenchmarks} and Ref.~\onlinecite{BuAnd2AgPolyenes} for both basis sets and approaches considered.

\begin{table}[t]
\centering
\caption{Excitation energies (eV) from MCCIPT2 with $c_{\text{min}}=5\times10^{-3}$ when using the def2-TZVP basis} \label{tbl:Def2TZVPExcitationEnergiesPT2}
\begin{tabular*}{8.5cm}{@{\extracolsep{\fill}}lccc}
\hline
\hline
Molecule & Excitation &  MCCIPT2   & PT2 configurations \\
\hline

Formaldehyde & 1 $^{1}A_{2}$  & 4.33   &  $1.2\times 10^{7}$    \\
 & 1 $^{1}B_{1}$ & 8.87  &    $1.3 \times 10^{7}$ \\
 & 2 $^{1}A_{1}$   &  10.04 &  $1.5 \times 10^{7}$\\

 Ethene & 1 $^{1}B_{1u}$  & 8.03    & $1.1 \times 10^{7}$  \\

Cyclopropene & 1 $^{1}B_{1}$  & 7.53   & $1.1\times 10^{8}$ \\
 & 1 $^{1}B_{2}$  & 6.20  & $1.4\times 10^{7}$\\

Formamide& 1 $^{1}A''$  & 6.49  & $1.1 \times 10^{7}$ \\
 & 2 $^{1}A'$  & 8.56  &   $2.2\times10^{8}$ \\
 & 3 $^{1}A'$  & 9.33 &  $2.2\times10^{8}$ \\

E-Butadiene & 1 $^{1}B_{u}$ &  7.38  &  $6.7\times10^{7}$ \\
& 2 $^{1}A_{g}$ & 7.67  &  $3.1\times 10^{8}$ \\
\hline
\hline
\end{tabular*}
\end{table}

\section{Summary}
 We have demonstrated that state-averaged Monte Carlo configuration interaction (SA-MCCI) can be used to overcome problems arising from the removal of configurations important for the ground state in an MCCI calculation.  SA-MCCI was seen to be able to calculate conical intersections, avoided crossings, a dipole moment and vertical excitations.   The conical intersection due to symmetry of $H_{3}$ in the shape of an equilateral triangle was successfully calculated with SA-MCCI without using symmetry. Potential curves near the degeneracy and when considering moving one hydrogen around a circle centred on a vertex of the triangle were also accurately calculated.  We showed that the measure of the error between two potential curves ($\sigma_{\Delta E}$) that we introduced in Ref.~\onlinecite{MCCInatorb} could also be used as a fair way to measure the error when comparing approximate potential curves for multiple states  
with the FCI results.  The avoided crossing in the potential curve of the first two $^{1}A_{1}$ states of LiF was recovered when using SA-MCCI in two bases and the potential curves had low errors when compared with the FCI results despite only using a very small fraction of the FCI space: a few thousand CSFs were needed compared with FCI spaces of around $10^{7}$ SDs.  The seam of conical intersections for the $1$ $^{3} A_{2}$ and $2$ $^{3}B_{1}$ states in CH\subscript{2} was fairly well reproduced by SA-MCCI although there was a slight deviation from degeneracy for some values.  When using geometries corresponding to $C_{s}$ symmetry and a smaller basis then potential curves for the first three triplet states of CH\subscript{2} agreed well with the FCI results and the use of CSFs in MCCI allowed a FCI state not present in the SA-MCCI triplet calculation to be revealed as a quintet. We saw that SA-MCCI can also be useful in the calculation of properties of excited states: the oscillations in the MCCI dipole of the $2$ $^{1}A_{1}$ state of carbon monoxide were eliminated by using SA-MCCI.

   We then considered the excitation energies of a collection of organic molecules up to the size of butadiene and compared SA-MCCI results with those of CASPT2 from Ref.~\onlinecite{OrganicBenchmarks} when using the def1-TZVP basis. We found that in general the excitation energies from SA-MCCI were larger than the CASPT2 results of Ref.~\onlinecite{OrganicBenchmarks} and the difference tends to be greater when the FCI space is much larger, while decreasing the cut-off lowers the MCCI excitation energy. The ordering of the states was the same.  The results agreed reasonably well for formaldehyde, ethene and cyclopropene in that SA-MCCI and CASPT2 were within 1 eV of each other.  For formamide the difference was more than 1 eV and for e-butadiene it was as large as 3 eV suggesting that the FCI space is too large here for MCCI to describe sufficiently well at a cut-off of $c_{\text{min}}=5\times10^{-4}$. The agreement in oscillator strength values was reasonable except for the $2$ $^{1} A_{1}$ state of formamide.  We then looked at using MCCIPT2 and approximate natural orbitals albeit with a larger cut-off of $5\times 10^{-3}$ for the SA-MCCI calculation so that the MCCIPT2 calculation would not have to consider more than a billion CSFs.  The results were now in better agreement with those of CASPT2 with a difference of always less than 1.1 eV with formamide having the largest discrepancy.  Interestingly the order of the two fairly similar cyclopropene excitations in MCCIPT2 was different to the CASPT2 results.  This remained the same in MCCIPT2 when the cutoff was lowered to $c_{\text{min}}=3\times10^{-3}$ although the energies were now closer.  We also used MCCIPT2 with the def2-TZVP basis for the excitation energies of these systems.  The only notable difference was that the 1 $^{1}B_{u}$ excitation of e-butadiene that was about 1.5eV higher than in the def1-TZVP basis suggesting that more configurations needed to be considered from the very large FCI space ($10^{28}$) of this system to get a more stable value.  This could be achieved by improving the efficiency of MCCIPT2, perhaps through parallelisation of the PT2 duplicate removal, to allow smaller cut-offs to be used for these very large FCI spaces. 

\acknowledgements{We thank the European Research Council (ERC) for funding under the European Union's Seventh Framework Programme (FP7/2007-2013)/ERC Grant No. 258990.}  

\providecommand{\noopsort}[1]{}\providecommand{\singleletter}[1]{#1}%


\begin{thebibliography}{31}%
\makeatletter
\providecommand \@ifxundefined [1]{%
 \@ifx{#1\undefined}
}%
\providecommand \@ifnum [1]{%
 \ifnum #1\expandafter \@firstoftwo
 \else \expandafter \@secondoftwo
 \fi
}%
\providecommand \@ifx [1]{%
 \ifx #1\expandafter \@firstoftwo
 \else \expandafter \@secondoftwo
 \fi
}%
\providecommand \natexlab [1]{#1}%
\providecommand \enquote  [1]{``#1''}%
\providecommand \bibnamefont  [1]{#1}%
\providecommand \bibfnamefont [1]{#1}%
\providecommand \citenamefont [1]{#1}%
\providecommand \href@noop [0]{\@secondoftwo}%
\providecommand \href [0]{\begingroup \@sanitize@url \@href}%
\providecommand \@href[1]{\@@startlink{#1}\@@href}%
\providecommand \@@href[1]{\endgroup#1\@@endlink}%
\providecommand \@sanitize@url [0]{\catcode `\\12\catcode `\$12\catcode
  `\&12\catcode `\#12\catcode `\^12\catcode `\_12\catcode `\%12\relax}%
\providecommand \@@startlink[1]{}%
\providecommand \@@endlink[0]{}%
\providecommand \url  [0]{\begingroup\@sanitize@url \@url }%
\providecommand \@url [1]{\endgroup\@href {#1}{\urlprefix }}%
\providecommand \urlprefix  [0]{URL }%
\providecommand \Eprint [0]{\href }%
\@ifxundefined \urlstyle {%
  \providecommand \doi  [0]{\begingroup \@sanitize@url \@doi}%
  \providecommand \@doi [1]{\endgroup \@@startlink {\doibase
  #1}doi:\discretionary {}{}{}#1\@@endlink }%
}{%
  \providecommand \doi  [0]{doi:\discretionary{}{}{}\begingroup
  \urlstyle{rm}\Url }%
}%
\providecommand \doibase [0]{http://dx.doi.org/}%
\providecommand \Doi [0]{\begingroup \@sanitize@url \@Doi }%
\providecommand \@Doi  [1]{\endgroup\@@startlink{\doibase#1}\@@Doi}%
\providecommand \@@Doi [1]{#1\@@endlink}%
\providecommand \selectlanguage [0]{\@gobble}%
\providecommand \bibinfo  [0]{\@secondoftwo}%
\providecommand \bibfield  [0]{\@secondoftwo}%
\providecommand \translation [1]{[#1]}%
\providecommand \BibitemOpen [0]{}%
\providecommand \bibitemStop [0]{}%
\providecommand \bibitemNoStop [0]{.\EOS\space}%
\providecommand \EOS [0]{\spacefactor3000\relax}%
\providecommand \BibitemShut  [1]{\csname bibitem#1\endcsname}%
\bibitem [{\citenamefont {Greer}(1998)}]{mcciGreer98}%
  \BibitemOpen
  \bibfield  {author} {\bibinfo {author} {\bibfnamefont {J.~C.}\ \bibnamefont
  {Greer}},\ }\href@noop {} {\bibfield  {journal} {\bibinfo  {journal} {J.
  Comp. Phys.},\ }\textbf {\bibinfo {volume} {146}},\ \bibinfo {pages} {181}
  (\bibinfo {year} {1998})}\BibitemShut {NoStop}%
\bibitem [{\citenamefont {Tong}\ \emph {et~al.}(2000)\citenamefont {Tong},
  \citenamefont {Nolan}, \citenamefont {Cheng},\ and\ \citenamefont
  {Greer}}]{mccicodeGreer}%
  \BibitemOpen
  \bibfield  {author} {\bibinfo {author} {\bibfnamefont {L.}~\bibnamefont
  {Tong}}, \bibinfo {author} {\bibfnamefont {M.}~\bibnamefont {Nolan}},
  \bibinfo {author} {\bibfnamefont {T.}~\bibnamefont {Cheng}}, \ and\ \bibinfo
  {author} {\bibfnamefont {J.~C.}\ \bibnamefont {Greer}},\ }\href@noop {}
  {\bibfield  {journal} {\bibinfo  {journal} {Comp. Phys. Comm.},\ }\textbf
  {\bibinfo {volume} {131}},\ \bibinfo {pages} {142} (\bibinfo {year}
  {2000})}\BibitemShut {NoStop}%
\bibitem [{\citenamefont {Gy\H{o}rffy}\ \emph {et~al.}(2008)\citenamefont
  {Gy\H{o}rffy}, \citenamefont {Bartlett},\ and\ \citenamefont
  {Greer}}]{GreerMcciSpectra}%
  \BibitemOpen
  \bibfield  {author} {\bibinfo {author} {\bibfnamefont {W.}~\bibnamefont
  {Gy\H{o}rffy}}, \bibinfo {author} {\bibfnamefont {R.~J.}\ \bibnamefont
  {Bartlett}}, \ and\ \bibinfo {author} {\bibfnamefont {J.~C.}\ \bibnamefont
  {Greer}},\ }\href@noop {} {\bibfield  {journal} {\bibinfo  {journal} {J.
  Chem. Phys.},\ }\textbf {\bibinfo {volume} {129}},\ \bibinfo {pages} {064103}
  (\bibinfo {year} {2008})}\BibitemShut {NoStop}%
\bibitem [{\citenamefont {Coe}\ \emph {et~al.}(2012)\citenamefont {Coe},
  \citenamefont {Taylor},\ and\ \citenamefont {Paterson}}]{MCCIpotentials}%
  \BibitemOpen
  \bibfield  {author} {\bibinfo {author} {\bibfnamefont {J.~P.}\ \bibnamefont
  {Coe}}, \bibinfo {author} {\bibfnamefont {D.~J.}\ \bibnamefont {Taylor}}, \
  and\ \bibinfo {author} {\bibfnamefont {M.~J.}\ \bibnamefont {Paterson}},\
  }\href@noop {} {\bibfield  {journal} {\bibinfo  {journal} {J. Chem. Phys.},\
  }\textbf {\bibinfo {volume} {137}},\ \bibinfo {pages} {194111} (\bibinfo
  {year} {2012})}\BibitemShut {NoStop}%
\bibitem [{\citenamefont {Coe}\ \emph {et~al.}(2013)\citenamefont {Coe},
  \citenamefont {Taylor},\ and\ \citenamefont {Paterson}}]{MCCIdipoles}%
  \BibitemOpen
  \bibfield  {author} {\bibinfo {author} {\bibfnamefont {J.~P.}\ \bibnamefont
  {Coe}}, \bibinfo {author} {\bibfnamefont {D.~J.}\ \bibnamefont {Taylor}}, \
  and\ \bibinfo {author} {\bibfnamefont {M.~J.}\ \bibnamefont {Paterson}},\
  }\href@noop {} {\bibfield  {journal} {\bibinfo  {journal} {J. Comput.
  Chem.},\ }\textbf {\bibinfo {volume} {34}},\ \bibinfo {pages} {1083}
  (\bibinfo {year} {2013})}\BibitemShut {NoStop}%
\bibitem [{\citenamefont {Stanton}\ and\ \citenamefont
  {Bartlett}(1993)}]{EOMCCoverview}%
  \BibitemOpen
  \bibfield  {author} {\bibinfo {author} {\bibfnamefont {J.~F.}\ \bibnamefont
  {Stanton}}\ and\ \bibinfo {author} {\bibfnamefont {R.~J.}\ \bibnamefont
  {Bartlett}},\ }\href@noop {} {\bibfield  {journal} {\bibinfo  {journal} {J.
  Chem. Phys.},\ }\textbf {\bibinfo {volume} {98}},\ \bibinfo {pages} {7029}
  (\bibinfo {year} {1993})}\BibitemShut {NoStop}%
\bibitem [{\citenamefont {Siegbahn}\ \emph {et~al.}(1981)\citenamefont
  {Siegbahn}, \citenamefont {Alml\"{o}f}, \citenamefont {Heiberg},\ and\
  \citenamefont {Roos}}]{siegbahn:2384}%
  \BibitemOpen
  \bibfield  {author} {\bibinfo {author} {\bibfnamefont {P.~E.~M.}\
  \bibnamefont {Siegbahn}}, \bibinfo {author} {\bibfnamefont {J.}~\bibnamefont
  {Alml\"{o}f}}, \bibinfo {author} {\bibfnamefont {A.}~\bibnamefont {Heiberg}},
  \ and\ \bibinfo {author} {\bibfnamefont {B.~O.}\ \bibnamefont {Roos}},\
  }\href@noop {} {\bibfield  {journal} {\bibinfo  {journal} {J. Chem. Phys.},\
  }\textbf {\bibinfo {volume} {74}},\ \bibinfo {pages} {2384} (\bibinfo {year}
  {1981})}\BibitemShut {NoStop}%
\bibitem [{\citenamefont {Andersson}\ \emph {et~al.}(1990)\citenamefont
  {Andersson}, \citenamefont {Malmqvist}, \citenamefont {Roos}, \citenamefont
  {Sadlej},\ and\ \citenamefont {Wolinski}}]{CASPT2}%
  \BibitemOpen
  \bibfield  {author} {\bibinfo {author} {\bibfnamefont {K.}~\bibnamefont
  {Andersson}}, \bibinfo {author} {\bibfnamefont {P.-{\AA}.}\ \bibnamefont
  {Malmqvist}}, \bibinfo {author} {\bibfnamefont {B.~O.}\ \bibnamefont {Roos}},
  \bibinfo {author} {\bibfnamefont {A.~J.}\ \bibnamefont {Sadlej}}, \ and\
  \bibinfo {author} {\bibfnamefont {K.}~\bibnamefont {Wolinski}},\ }\href@noop
  {} {\bibfield  {journal} {\bibinfo  {journal} {J. Phys. Chem.},\ }\textbf
  {\bibinfo {volume} {94}},\ \bibinfo {pages} {5483} (\bibinfo {year}
  {1990})}\BibitemShut {NoStop}%
\bibitem [{\citenamefont {Ohtsuka}\ and\ \citenamefont
  {Nagase}(2010)}]{Nagase10PMC}%
  \BibitemOpen
  \bibfield  {author} {\bibinfo {author} {\bibfnamefont {Y.}~\bibnamefont
  {Ohtsuka}}\ and\ \bibinfo {author} {\bibfnamefont {S.}~\bibnamefont
  {Nagase}},\ }\href@noop {} {\bibfield  {journal} {\bibinfo  {journal} {Chem.
  Phys. Lett.},\ }\textbf {\bibinfo {volume} {485}},\ \bibinfo {pages} {367}
  (\bibinfo {year} {2010})}\BibitemShut {NoStop}%
\bibitem [{\citenamefont {Booth}\ and\ \citenamefont
  {Chan}(2012)}]{FCIQMCexcited}%
  \BibitemOpen
  \bibfield  {author} {\bibinfo {author} {\bibfnamefont {G.~H.}\ \bibnamefont
  {Booth}}\ and\ \bibinfo {author} {\bibfnamefont {G.~K.-L.}\ \bibnamefont
  {Chan}},\ }\href@noop {} {\bibfield  {journal} {\bibinfo  {journal} {J. Chem.
  Phys.},\ }\textbf {\bibinfo {volume} {137}},\ \bibinfo {pages} {191102}
  (\bibinfo {year} {2012})}\BibitemShut {NoStop}%
\bibitem [{\citenamefont {Ten-no}(2013)}]{MSQMC}%
  \BibitemOpen
  \bibfield  {author} {\bibinfo {author} {\bibfnamefont {S.}~\bibnamefont
  {Ten-no}},\ }\href@noop {} {\bibfield  {journal} {\bibinfo  {journal} {J.
  Chem. Phys.},\ }\textbf {\bibinfo {volume} {138}},\ \bibinfo {pages} {164126}
  (\bibinfo {year} {2013})}\BibitemShut {NoStop}%
\bibitem [{\citenamefont {Coe}\ and\ \citenamefont
  {Paterson}(2012)}]{MCCInatorb}%
  \BibitemOpen
  \bibfield  {author} {\bibinfo {author} {\bibfnamefont {J.~P.}\ \bibnamefont
  {Coe}}\ and\ \bibinfo {author} {\bibfnamefont {M.~J.}\ \bibnamefont
  {Paterson}},\ }\href@noop {} {\bibfield  {journal} {\bibinfo  {journal} {J.
  Chem. Phys.},\ }\textbf {\bibinfo {volume} {137}},\ \bibinfo {pages} {204108}
  (\bibinfo {year} {2012})}\BibitemShut {NoStop}%
\bibitem [{\citenamefont {Werner}\ \emph {et~al.}(2010)\citenamefont {Werner},
  \citenamefont {Knowles}, \citenamefont {Knizia}, \citenamefont {Manby},
  \citenamefont {{Sch\"{u}tz}}, \citenamefont {Celani}, \citenamefont {Korona},
  \citenamefont {Lindh}, \citenamefont {Mitrushenkov}, \citenamefont {Rauhut},
  \citenamefont {Shamasundar}, \citenamefont {Adler}, \citenamefont {Amos},
  \citenamefont {Bernhardsson}, \citenamefont {Berning}, \citenamefont
  {Cooper}, \citenamefont {Deegan}, \citenamefont {Dobbyn}, \citenamefont
  {Eckert}, \citenamefont {Goll}, \citenamefont {Hampel}, \citenamefont
  {Hesselmann}, \citenamefont {Hetzer}, \citenamefont {Hrenar}, \citenamefont
  {Jansen}, \citenamefont {K\"oppl}, \citenamefont {Liu}, \citenamefont
  {Lloyd}, \citenamefont {Mata}, \citenamefont {May}, \citenamefont
  {McNicholas}, \citenamefont {Meyer}, \citenamefont {Mura}, \citenamefont
  {Nicklass}, \citenamefont {O'Neill}, \citenamefont {Palmieri}, \citenamefont
  {Pfl\"uger}, \citenamefont {Pitzer}, \citenamefont {Reiher}, \citenamefont
  {Shiozaki}, \citenamefont {Stoll}, \citenamefont {Stone}, \citenamefont
  {Tarroni}, \citenamefont {Thorsteinsson}, \citenamefont {Wang},\ and\
  \citenamefont {Wolf}}]{Molpro}%
  \BibitemOpen
  \bibfield  {author} {\bibinfo {author} {\bibfnamefont {H.-J.}\ \bibnamefont
  {Werner}}, \bibinfo {author} {\bibfnamefont {P.~J.}\ \bibnamefont {Knowles}},
  \bibinfo {author} {\bibfnamefont {G.}~\bibnamefont {Knizia}}, \bibinfo
  {author} {\bibfnamefont {F.~R.}\ \bibnamefont {Manby}}, \bibinfo {author}
  {\bibfnamefont {M.}~\bibnamefont {{Sch\"{u}tz}}}, \bibinfo {author}
  {\bibfnamefont {P.}~\bibnamefont {Celani}}, \bibinfo {author} {\bibfnamefont
  {T.}~\bibnamefont {Korona}}, \bibinfo {author} {\bibfnamefont
  {R.}~\bibnamefont {Lindh}}, \bibinfo {author} {\bibfnamefont
  {A.}~\bibnamefont {Mitrushenkov}}, \bibinfo {author} {\bibfnamefont
  {G.}~\bibnamefont {Rauhut}}, \bibinfo {author} {\bibfnamefont {K.~R.}\
  \bibnamefont {Shamasundar}}, \bibinfo {author} {\bibfnamefont {T.~B.}\
  \bibnamefont {Adler}}, \bibinfo {author} {\bibfnamefont {R.~D.}\ \bibnamefont
  {Amos}}, \bibinfo {author} {\bibfnamefont {A.}~\bibnamefont {Bernhardsson}},
  \bibinfo {author} {\bibfnamefont {A.}~\bibnamefont {Berning}}, \bibinfo
  {author} {\bibfnamefont {D.~L.}\ \bibnamefont {Cooper}}, \bibinfo {author}
  {\bibfnamefont {M.~J.~O.}\ \bibnamefont {Deegan}}, \bibinfo {author}
  {\bibfnamefont {A.~J.}\ \bibnamefont {Dobbyn}}, \bibinfo {author}
  {\bibfnamefont {F.}~\bibnamefont {Eckert}}, \bibinfo {author} {\bibfnamefont
  {E.}~\bibnamefont {Goll}}, \bibinfo {author} {\bibfnamefont {C.}~\bibnamefont
  {Hampel}}, \bibinfo {author} {\bibfnamefont {A.}~\bibnamefont {Hesselmann}},
  \bibinfo {author} {\bibfnamefont {G.}~\bibnamefont {Hetzer}}, \bibinfo
  {author} {\bibfnamefont {T.}~\bibnamefont {Hrenar}}, \bibinfo {author}
  {\bibfnamefont {G.}~\bibnamefont {Jansen}}, \bibinfo {author} {\bibfnamefont
  {C.}~\bibnamefont {K\"oppl}}, \bibinfo {author} {\bibfnamefont
  {Y.}~\bibnamefont {Liu}}, \bibinfo {author} {\bibfnamefont {A.~W.}\
  \bibnamefont {Lloyd}}, \bibinfo {author} {\bibfnamefont {R.~A.}\ \bibnamefont
  {Mata}}, \bibinfo {author} {\bibfnamefont {A.~J.}\ \bibnamefont {May}},
  \bibinfo {author} {\bibfnamefont {S.~J.}\ \bibnamefont {McNicholas}},
  \bibinfo {author} {\bibfnamefont {W.}~\bibnamefont {Meyer}}, \bibinfo
  {author} {\bibfnamefont {M.~E.}\ \bibnamefont {Mura}}, \bibinfo {author}
  {\bibfnamefont {A.}~\bibnamefont {Nicklass}}, \bibinfo {author}
  {\bibfnamefont {D.~P.}\ \bibnamefont {O'Neill}}, \bibinfo {author}
  {\bibfnamefont {P.}~\bibnamefont {Palmieri}}, \bibinfo {author}
  {\bibfnamefont {K.}~\bibnamefont {Pfl\"uger}}, \bibinfo {author}
  {\bibfnamefont {R.}~\bibnamefont {Pitzer}}, \bibinfo {author} {\bibfnamefont
  {M.}~\bibnamefont {Reiher}}, \bibinfo {author} {\bibfnamefont
  {T.}~\bibnamefont {Shiozaki}}, \bibinfo {author} {\bibfnamefont
  {H.}~\bibnamefont {Stoll}}, \bibinfo {author} {\bibfnamefont {A.~J.}\
  \bibnamefont {Stone}}, \bibinfo {author} {\bibfnamefont {R.}~\bibnamefont
  {Tarroni}}, \bibinfo {author} {\bibfnamefont {T.}~\bibnamefont
  {Thorsteinsson}}, \bibinfo {author} {\bibfnamefont {M.}~\bibnamefont {Wang}},
  \ and\ \bibinfo {author} {\bibfnamefont {A.}~\bibnamefont {Wolf}},\
  }\href@noop {} {\enquote {\bibinfo {title} {Molpro, version 2010.1, a package
  of ab initio programs},}\ } (\bibinfo {year} {2010}),\ \bibinfo {note} {see
  http://www.molpro.net}\BibitemShut {NoStop}%
\bibitem [{\citenamefont {Lischka}\ \emph {et~al.}()\citenamefont {Lischka},
  \citenamefont {Shepard}, \citenamefont {Shavitt}, \citenamefont {Pitzer},
  \citenamefont {Dallos}, \citenamefont {Muller}, \citenamefont {Szalay},
  \citenamefont {Brown}, \citenamefont {Ahlrichs}, \citenamefont {Boehm},
  \citenamefont {Chang}, \citenamefont {Comeau}, \citenamefont {Gdanitz},
  \citenamefont {Dachsel}, \citenamefont {Ehrhardt}, \citenamefont {Ernzerhof},
  \citenamefont {Hochtl}, \citenamefont {Irle}, \citenamefont {Kedziora},
  \citenamefont {Kovar}, \citenamefont {Parasuk}, \citenamefont {Pepper},
  \citenamefont {Scharf}, \citenamefont {Schiffer}, \citenamefont {Schindler},
  \citenamefont {Schuler}, \citenamefont {Seth}, \citenamefont {Stahlberg},
  \citenamefont {Zhao}, \citenamefont {Yabushita}, \citenamefont {Zhang},
  \citenamefont {Barbatti}, \citenamefont {Matsika}, \citenamefont
  {Schuurmann}, \citenamefont {Yarkony}, \citenamefont {Brozell}, \citenamefont
  {Beck}, \citenamefont {Blaudeau}, \citenamefont {Ruckenbauer}, \citenamefont
  {Sellner}, \citenamefont {Plasser},\ and\ \citenamefont
  {Szymczak}}]{Columbus}%
  \BibitemOpen
  \bibfield  {author} {\bibinfo {author} {\bibfnamefont {H.}~\bibnamefont
  {Lischka}}, \bibinfo {author} {\bibfnamefont {R.}~\bibnamefont {Shepard}},
  \bibinfo {author} {\bibfnamefont {I.}~\bibnamefont {Shavitt}}, \bibinfo
  {author} {\bibfnamefont {R.~M.}\ \bibnamefont {Pitzer}}, \bibinfo {author}
  {\bibfnamefont {M.}~\bibnamefont {Dallos}}, \bibinfo {author} {\bibfnamefont
  {T.}~\bibnamefont {Muller}}, \bibinfo {author} {\bibfnamefont {P.~G.}\
  \bibnamefont {Szalay}}, \bibinfo {author} {\bibfnamefont {F.~B.}\
  \bibnamefont {Brown}}, \bibinfo {author} {\bibfnamefont {R.}~\bibnamefont
  {Ahlrichs}}, \bibinfo {author} {\bibfnamefont {H.~J.}\ \bibnamefont {Boehm}},
  \bibinfo {author} {\bibfnamefont {A.}~\bibnamefont {Chang}}, \bibinfo
  {author} {\bibfnamefont {D.~C.}\ \bibnamefont {Comeau}}, \bibinfo {author}
  {\bibfnamefont {R.}~\bibnamefont {Gdanitz}}, \bibinfo {author} {\bibfnamefont
  {H.}~\bibnamefont {Dachsel}}, \bibinfo {author} {\bibfnamefont
  {C.}~\bibnamefont {Ehrhardt}}, \bibinfo {author} {\bibfnamefont
  {M.}~\bibnamefont {Ernzerhof}}, \bibinfo {author} {\bibfnamefont
  {P.}~\bibnamefont {Hochtl}}, \bibinfo {author} {\bibfnamefont
  {S.}~\bibnamefont {Irle}}, \bibinfo {author} {\bibfnamefont {G.}~\bibnamefont
  {Kedziora}}, \bibinfo {author} {\bibfnamefont {T.}~\bibnamefont {Kovar}},
  \bibinfo {author} {\bibfnamefont {V.}~\bibnamefont {Parasuk}}, \bibinfo
  {author} {\bibfnamefont {M.~J.~M.}\ \bibnamefont {Pepper}}, \bibinfo {author}
  {\bibfnamefont {P.}~\bibnamefont {Scharf}}, \bibinfo {author} {\bibfnamefont
  {H.}~\bibnamefont {Schiffer}}, \bibinfo {author} {\bibfnamefont
  {M.}~\bibnamefont {Schindler}}, \bibinfo {author} {\bibfnamefont
  {M.}~\bibnamefont {Schuler}}, \bibinfo {author} {\bibfnamefont
  {M.}~\bibnamefont {Seth}}, \bibinfo {author} {\bibfnamefont {E.~A.}\
  \bibnamefont {Stahlberg}}, \bibinfo {author} {\bibfnamefont {J.-G.}\
  \bibnamefont {Zhao}}, \bibinfo {author} {\bibfnamefont {S.}~\bibnamefont
  {Yabushita}}, \bibinfo {author} {\bibfnamefont {Z.}~\bibnamefont {Zhang}},
  \bibinfo {author} {\bibfnamefont {M.}~\bibnamefont {Barbatti}}, \bibinfo
  {author} {\bibfnamefont {S.}~\bibnamefont {Matsika}}, \bibinfo {author}
  {\bibfnamefont {M.}~\bibnamefont {Schuurmann}}, \bibinfo {author}
  {\bibfnamefont {D.~R.}\ \bibnamefont {Yarkony}}, \bibinfo {author}
  {\bibfnamefont {S.~R.}\ \bibnamefont {Brozell}}, \bibinfo {author}
  {\bibfnamefont {E.~V.}\ \bibnamefont {Beck}}, \bibinfo {author}
  {\bibfnamefont {J.-P.}\ \bibnamefont {Blaudeau}}, \bibinfo {author}
  {\bibfnamefont {M.}~\bibnamefont {Ruckenbauer}}, \bibinfo {author}
  {\bibfnamefont {B.}~\bibnamefont {Sellner}}, \bibinfo {author} {\bibfnamefont
  {F.}~\bibnamefont {Plasser}}, \ and\ \bibinfo {author} {\bibfnamefont
  {J.~J.}\ \bibnamefont {Szymczak}},\ }\href@noop {} {\enquote {\bibinfo
  {title} {Columbus, an ab initio electronic structure program, release 5.9.2,
  http://www.univie.ac.at/columbus (2008)},}\ }\BibitemShut {NoStop}%
\bibitem [{\citenamefont {Gy\H{o}rffy}(2007)}]{GyorffyThesis}%
  \BibitemOpen
  \bibfield  {author} {\bibinfo {author} {\bibfnamefont {W.}~\bibnamefont
  {Gy\H{o}rffy}},\ }\emph {\bibinfo {title} {Monte Carlo Configuration
  Interaction Method for Calculation of Electronic Spectra of Molecules}},\
  \href@noop {} {Ph.D. thesis},\ \bibinfo  {school} {University College Cork}
  (\bibinfo {year} {2007})\BibitemShut {NoStop}%
\bibitem [{\citenamefont {Liu}(1979)}]{LiuReportNumerical}%
  \BibitemOpen
  \bibfield  {author} {\bibinfo {author} {\bibfnamefont {B.}~\bibnamefont
  {Liu}},\ }in\ \href@noop {} {\emph {\bibinfo {booktitle} {Report on the
  workshop numerical algorithms in chemistry: algebraic methods}}},\ \bibinfo
  {editor} {edited by\ \bibinfo {editor} {\bibfnamefont {C.}~\bibnamefont
  {Moler}}\ and\ \bibinfo {editor} {\bibfnamefont {I.}~\bibnamefont
  {Shavitt}}}\ (\bibinfo {year} {1979})\ p.~\bibinfo {pages} {49}\BibitemShut
  {NoStop}%
\bibitem [{\citenamefont {Longuet-Higgins}(1975)}]{LHphase}%
  \BibitemOpen
  \bibfield  {author} {\bibinfo {author} {\bibfnamefont {H.}~\bibnamefont
  {Longuet-Higgins}},\ }\href@noop {} {\bibfield  {journal} {\bibinfo
  {journal} {Proc. R. Soc. London A},\ }\textbf {\bibinfo {volume} {344}},\
  \bibinfo {pages} {147} (\bibinfo {year} {1975})}\BibitemShut {NoStop}%
\bibitem [{\citenamefont {L\"{o}wdin}(1955)}]{Lowdin55}%
  \BibitemOpen
  \bibfield  {author} {\bibinfo {author} {\bibfnamefont {P.-O.}\ \bibnamefont
  {L\"{o}wdin}},\ }\href@noop {} {\bibfield  {journal} {\bibinfo  {journal}
  {Phys. Rev.},\ }\textbf {\bibinfo {volume} {97}},\ \bibinfo {pages} {1474}
  (\bibinfo {year} {1955})}\BibitemShut {NoStop}%
\bibitem [{\citenamefont {{Bauschlicher Jr.}}\ and\ \citenamefont
  {Langhoff}(1988)}]{Bauschlicher88}%
  \BibitemOpen
  \bibfield  {author} {\bibinfo {author} {\bibfnamefont {C.~W.}\ \bibnamefont
  {{Bauschlicher Jr.}}}\ and\ \bibinfo {author} {\bibfnamefont {S.~R.}\
  \bibnamefont {Langhoff}},\ }\href@noop {} {\bibfield  {journal} {\bibinfo
  {journal} {J. Chem. Phys.},\ }\textbf {\bibinfo {volume} {89}},\ \bibinfo
  {pages} {4246} (\bibinfo {year} {1988})}\BibitemShut {NoStop}%
\bibitem [{\citenamefont {Malrieu}\ \emph {et~al.}(1995)\citenamefont
  {Malrieu}, \citenamefont {Heully},\ and\ \citenamefont
  {Zaitzevskii}}]{Malrieu95}%
  \BibitemOpen
  \bibfield  {author} {\bibinfo {author} {\bibfnamefont {J.-P.}\ \bibnamefont
  {Malrieu}}, \bibinfo {author} {\bibfnamefont {J.-L.}\ \bibnamefont {Heully}},
  \ and\ \bibinfo {author} {\bibfnamefont {A.}~\bibnamefont {Zaitzevskii}},\
  }\href@noop {} {\bibfield  {journal} {\bibinfo  {journal} {Theor. Chim.
  Acta},\ }\textbf {\bibinfo {volume} {90}},\ \bibinfo {pages} {167} (\bibinfo
  {year} {1995})}\BibitemShut {NoStop}%
\bibitem [{\citenamefont {Finley}\ \emph {et~al.}(1998)\citenamefont {Finley},
  \citenamefont {Malmqvist}, \citenamefont {Roos},\ and\ \citenamefont
  {Serrano-Andr\'{e}s}}]{Finley98}%
  \BibitemOpen
  \bibfield  {author} {\bibinfo {author} {\bibfnamefont {J.}~\bibnamefont
  {Finley}}, \bibinfo {author} {\bibfnamefont {P.-{\AA}.}\ \bibnamefont
  {Malmqvist}}, \bibinfo {author} {\bibfnamefont {B.~O.}\ \bibnamefont {Roos}},
  \ and\ \bibinfo {author} {\bibfnamefont {L.}~\bibnamefont
  {Serrano-Andr\'{e}s}},\ }\href@noop {} {\bibfield  {journal} {\bibinfo
  {journal} {Chem. Phys. Lett.},\ }\textbf {\bibinfo {volume} {288}},\ \bibinfo
  {pages} {299} (\bibinfo {year} {1998})}\BibitemShut {NoStop}%
\bibitem [{\citenamefont {Varandas}(2009)}]{VarandasLiF09}%
  \BibitemOpen
  \bibfield  {author} {\bibinfo {author} {\bibfnamefont {A.~J.~C.}\
  \bibnamefont {Varandas}},\ }\href@noop {} {\bibfield  {journal} {\bibinfo
  {journal} {J. Chem. Phys.},\ }\textbf {\bibinfo {volume} {131}},\ \bibinfo
  {pages} {124128} (\bibinfo {year} {2009})}\BibitemShut {NoStop}%
\bibitem [{\citenamefont {Greer}(1995)}]{mcciGreer95}%
  \BibitemOpen
  \bibfield  {author} {\bibinfo {author} {\bibfnamefont {J.~C.}\ \bibnamefont
  {Greer}},\ }\href@noop {} {\bibfield  {journal} {\bibinfo  {journal} {J.
  Chem. Phys.},\ }\textbf {\bibinfo {volume} {103}},\ \bibinfo {pages} {1821}
  (\bibinfo {year} {1995})}\BibitemShut {NoStop}%
\bibitem [{\citenamefont {Knowles}\ and\ \citenamefont
  {Handy}(1984)}]{MolproFCI1}%
  \BibitemOpen
  \bibfield  {author} {\bibinfo {author} {\bibfnamefont {P.~J.}\ \bibnamefont
  {Knowles}}\ and\ \bibinfo {author} {\bibfnamefont {N.~C.}\ \bibnamefont
  {Handy}},\ }\href@noop {} {\bibfield  {journal} {\bibinfo  {journal} {Chem.
  Phys. Lett.},\ }\textbf {\bibinfo {volume} {111}},\ \bibinfo {pages} {315}
  (\bibinfo {year} {1984})}\BibitemShut {NoStop}%
\bibitem [{\citenamefont {Yarkony}(1996)}]{YarkonyCH2}%
  \BibitemOpen
  \bibfield  {author} {\bibinfo {author} {\bibfnamefont {D.~R.}\ \bibnamefont
  {Yarkony}},\ }\href@noop {} {\bibfield  {journal} {\bibinfo  {journal} {J.
  Chem. Phys.},\ }\textbf {\bibinfo {volume} {104}},\ \bibinfo {pages} {2932}
  (\bibinfo {year} {1996})}\BibitemShut {NoStop}%
\bibitem [{\citenamefont {Schreiber}\ \emph {et~al.}(2008)\citenamefont
  {Schreiber}, \citenamefont {Silva-Junior}, \citenamefont {Sauer},\ and\
  \citenamefont {Thiel}}]{OrganicBenchmarks}%
  \BibitemOpen
  \bibfield  {author} {\bibinfo {author} {\bibfnamefont {M.}~\bibnamefont
  {Schreiber}}, \bibinfo {author} {\bibfnamefont {M.~R.}\ \bibnamefont
  {Silva-Junior}}, \bibinfo {author} {\bibfnamefont {S.~P.~A.}\ \bibnamefont
  {Sauer}}, \ and\ \bibinfo {author} {\bibfnamefont {W.}~\bibnamefont
  {Thiel}},\ }\href@noop {} {\bibfield  {journal} {\bibinfo  {journal} {J.
  Chem. Phys.},\ }\textbf {\bibinfo {volume} {128}},\ \bibinfo {pages} {134110}
  (\bibinfo {year} {2008})}\BibitemShut {NoStop}%
\bibitem [{\citenamefont {Sch\"{a}fer}\ \emph {et~al.}(1992)\citenamefont
  {Sch\"{a}fer}, \citenamefont {Horn},\ and\ \citenamefont
  {Ahlrichs}}]{def1TZVPbasisRef}%
  \BibitemOpen
  \bibfield  {author} {\bibinfo {author} {\bibfnamefont {A.}~\bibnamefont
  {Sch\"{a}fer}}, \bibinfo {author} {\bibfnamefont {H.}~\bibnamefont {Horn}}, \
  and\ \bibinfo {author} {\bibfnamefont {R.}~\bibnamefont {Ahlrichs}},\
  }\href@noop {} {\bibfield  {journal} {\bibinfo  {journal} {J. Chem. Phys.},\
  }\textbf {\bibinfo {volume} {97}},\ \bibinfo {pages} {2571} (\bibinfo {year}
  {1992})}\BibitemShut {NoStop}%
\bibitem [{\citenamefont {Nakayama}\ \emph {et~al.}(1998)\citenamefont
  {Nakayama}, \citenamefont {Nakano},\ and\ \citenamefont
  {Hirao}}]{BuAnd2AgPolyenes}%
  \BibitemOpen
  \bibfield  {author} {\bibinfo {author} {\bibfnamefont {K.}~\bibnamefont
  {Nakayama}}, \bibinfo {author} {\bibfnamefont {H.}~\bibnamefont {Nakano}}, \
  and\ \bibinfo {author} {\bibfnamefont {K.}~\bibnamefont {Hirao}},\
  }\href@noop {} {\bibfield  {journal} {\bibinfo  {journal} {Int. J. Quantum
  Chem.},\ }\textbf {\bibinfo {volume} {66}},\ \bibinfo {pages} {157} (\bibinfo
  {year} {1998})}\BibitemShut {NoStop}%
\bibitem [{\citenamefont {Harrison}(1991)}]{HarrisonFCIperturbation}%
  \BibitemOpen
  \bibfield  {author} {\bibinfo {author} {\bibfnamefont {R.~J.}\ \bibnamefont
  {Harrison}},\ }\href@noop {} {\bibfield  {journal} {\bibinfo  {journal} {J.
  Chem. Phys.},\ }\textbf {\bibinfo {volume} {94}},\ \bibinfo {pages} {5021}
  (\bibinfo {year} {1991})}\BibitemShut {NoStop}%
\bibitem [{\citenamefont {Pople}\ \emph {et~al.}(1987)\citenamefont {Pople},
  \citenamefont {Head-Gordon},\ and\ \citenamefont {Raghavachari}}]{QCISD}%
  \BibitemOpen
  \bibfield  {author} {\bibinfo {author} {\bibfnamefont {J.~A.}\ \bibnamefont
  {Pople}}, \bibinfo {author} {\bibfnamefont {M.}~\bibnamefont {Head-Gordon}},
  \ and\ \bibinfo {author} {\bibfnamefont {K.}~\bibnamefont {Raghavachari}},\
  }\href@noop {} {\bibfield  {journal} {\bibinfo  {journal} {J. Chem. Phys.},\
  }\textbf {\bibinfo {volume} {87}},\ \bibinfo {pages} {5968} (\bibinfo {year}
  {1987})}\BibitemShut {NoStop}%
\bibitem [{\citenamefont {Weigend}\ and\ \citenamefont
  {Ahlrichs}(2005)}]{def2TZVPbasisRef}%
  \BibitemOpen
  \bibfield  {author} {\bibinfo {author} {\bibfnamefont {F.}~\bibnamefont
  {Weigend}}\ and\ \bibinfo {author} {\bibfnamefont {R.}~\bibnamefont
  {Ahlrichs}},\ }\href@noop {} {\bibfield  {journal} {\bibinfo  {journal}
  {Phys. Chem. Chem. Phys.},\ }\textbf {\bibinfo {volume} {7}},\ \bibinfo
  {pages} {3297} (\bibinfo {year} {2005})}\BibitemShut {NoStop}%
\end{thebibliography}
\end{document}